\documentclass[review]{elsarticle}
\makeatletter
\def\ps@pprintTitle{%
 \let\@oddhead\@empty
 \let\@evenhead\@empty
 \def\@oddfoot{\centerline{\thepage}}%
 \let\@evenfoot\@oddfoot}
\makeatother

\pdfoutput=1
\usepackage{graphicx}
\usepackage{subfigure}
\usepackage{amsfonts, amsmath, amssymb}
\usepackage{mathabx}
\usepackage{booktabs,siunitx}
\usepackage[svgnames,table]{xcolor}
\usepackage[tableposition=above]{caption}
\usepackage{pifont}
\usepackage{bm}
\usepackage{cancel}
\usepackage{breqn}
\usepackage{caption}
\usepackage{color}
\usepackage{enumerate}
\usepackage{float}
\usepackage{hyperref}
\usepackage{soul}
\usepackage[english]{babel}
\usepackage[margin=1in]{geometry}
\usepackage{mathtools}
\usepackage{multirow}
\usepackage{setspace}
\usepackage{subfigure}
\usepackage{stmaryrd}
\usepackage{lineno}
\DeclareUnicodeCharacter{00B3}{\textsuperscript{3}}
\usepackage[ruled,linesnumbered]{algorithm2e}
\RestyleAlgo{ruled}
\SetKwComment{Comment}{/* }{ */}

\SetCommentSty{mycommfont}
\DeclareMathAlphabet{\mathpzc}{OT1}{pzc}{m}{it}

\newcommand \dd[2] {\frac{{\rm d} #1}{{\rm d} #2}}
\newcommand \dD[2] {\frac{{\rm d^2} #1}{{\rm d} {#2}^2}}
\renewcommand \d[1]{{\rm{d}} #1}
\newcommand \D [2]{\frac{\partial #1}{\partial #2}}

\renewcommand{\vec}[1]{\bm{\mathrm{#1}}}
\newcommand{\V}[1]{\bm{\mathrm{#1}}}

\def \grad{\nabla}

\def \q{\vec{q}}
\def \n{\vec{n}}
\def \s{\vec{s}}
\def \u{\vec{u}}

\def \Rzero{R_0}
\def \Rone{R_1}
\def \Rtwo{R_2}
\def \Rb{R_b}

\def \Rhatone{\widehat{R}_1}
\def \Rhatb{\widehat{R}_b}

\def \rhat{\widehat{r}}
\def \that{\widehat{t}}

\def \rtil{\widetilde{r}}
\def \ttil{\widetilde{t}}
\def \Rtilone{\widetilde{R}_1}
\def \Rtiltwo{\widetilde{R}_2}
\def \Rtilb{\widetilde{R}_b}

\def \rhos{\rho^{\rm S}}
\def \rhol{\rho^{\rm L}}
\def \rhov{\rho^{\rm V}}

\def \rrhols{\rho^{\rm LS}} 
\def \rrhosl{\rho^{\rm SL}} 
\def \rrhovl{\rho^{\rm VL}} 
\def \rrholv{\rho^{\rm LV}} 
\def \rrhosv{\rho^{\rm SV}}

\def \rrhosltil{\widetilde{\rho}^{\rm SL}}
\def \rrholvtil{\widetilde{\rho}^{\rm LV}}

\def \cps{C^{\rm S}}
\def \cpl{C^{\rm L}}
\def \cpv{C^{\rm V}}

\def \Ts{T^{\rm S}}
\def \Tl{T^{\rm L}}
\def \Tv{T^{\rm V}}
\def \Tr{T_r}
\def \Tinf{T_\infty}
\def \Tim{T_m^{\rm I}}
\def \Tiv{T_v^{\rm I}}

\def \DeltaT{\Delta T}

\def \ks{\kappa^{\rm S}}
\def \kl{\kappa^{\rm L}}
\def \kv{\kappa^{\rm V}}

\def \hs{h^{\rm S}}
\def \hl{h^{\rm L}}
\def \hv{h^{\rm V}}

\def \el{e^{\rm L}}
\def \es{e^{\rm S}}
\def \ev{e^{\rm V}}

\def \pl{p^{\rm L}}
\def \ps{p^{\rm S}}
\def \pv{p^{\rm V}}

\def \ul{u^{\rm L}}
\def \us{u^{\rm S}}
\def \uv{u^{\rm V}}

\def \uoneStar{u_1}

\def \Omegas{\Omega^{\rm S}}
\def \Omegal{\Omega^{\rm L}}
\def \Omegav{\Omega^{\rm V}}

\def \alphas{\alpha^{\rm S}}
\def \alphal{\alpha^{\rm L}}

\def \That{\widehat{T}}
\def \Thats{\widehat{T}^{\rm S}}
\def \Thatl{\widehat{T}^{\rm L}}
\def \Thatv{\widehat{T}^{\rm V}}
\def \Thatm{\widehat{T}_m}
\def \Thatinf{\widehat{T}_\infty}
\def \Thatim{\widehat{T}_m^{\rm I}}
\def \Thativ{\widehat{T}_v^{\rm I}}
\def \Ttilim{\widetilde{T}_m^{\rm I}}
\def \Ttiliv{\widetilde{T}_v^{\rm I}}
\def \Ttils{\widetilde{T}^{\rm S}}
\def \Ttill{\widetilde{T}^{\rm L}}
\def \Ttilv{\widetilde{T}^{\rm V}}

\def \uhat{\widehat{u}}
\def \uhats{\widehat{u}^{\rm S}}
\def \uhatl{\widehat{u}^{\rm L}}
\def \uhatv{\widehat{u}^{\rm V}}
\def \uhatone{\widehat{u}_1}

\def \betam{\beta_m}
\def \gammam{\gamma_m}
\def \deltam{\delta_m}
\def \Gammam{\Gamma_m}

\def \betav{\beta_v}
\def \bv{S_v}
\def \gammav{\gamma_v}
\def \deltav{\delta_v}
\def \Gammav{\Gamma_v}

\def \utwoStar{u_2}
\def \uhattwo{\widehat{u}_2}
\def \Rhattwo{\widehat{R}_2}
\def \Rhatb{\widehat{R}_b}

\def \sigmasl{\sigma_{\rm SL}} 
\def \sigmalv{\sigma_{\rm LV}}

\def \half{\frac{1}{2}}
\def \threehalf{\frac{3}{2}}

\def \Rmax{R_{\rm max}}
\def \Nmax{N_{\rm max}}
\def \Nmin{N_{\rm min}}

\graphicspath{{figures/}}

\usepackage{caption}  

\newcommand{\upperRomannumeral}[1]{\uppercase\expandafter{\romannumeral#1}}

\begin{document}
\let\today\relax
\let\underbrace\LaTeXunderbrace
\let\overbrace\LaTeXoverbrace

\begin{frontmatter}

\title{Fixed-grid sharp-interface numerical solutions to the three-phase spherical Stefan problem}
 \author[SDSU]{Yavkreet Swami}
 \author[SDSU]{Jacob Barajas}
\author[SDSU]{Amneet Pal Singh Bhalla\corref{mycorrespondingauthor}}
\ead{asbhalla@sdsu.edu}

\address[SDSU]{Department of Mechanical Engineering, San Diego State University, San Diego, CA}

\begin{abstract}
Many metal manufacturing processes involve phase change phenomena, which include melting, boiling, and vaporization. These phenomena often occur concurrently. A prototypical 1D model for understanding the phase change phenomena is the Stefan problem. There is a large body of literature discussing the analytical solution to the two-phase Stefan problem that describes only the melting or boiling of phase change materials (PCMs) with one moving interface. Density-change effects that induce additional fluid flow during phase change are generally neglected in the literature to simplify the math of the Stefan problem. In our recent work~\cite{Mehran2025analyticalmsnbc}, we provide analytical and numerical solutions to the three-phase Stefan problem with simultaneous occurrences of melting, solidification, boiling, and condensation in Cartesian coordinates. The problem was set in a semi-infinite domain with two moving interfaces---melt and boiling fronts. The solid, liquid, and vapor phases of the PCM have distinct thermophysical properties, including density, thermal conductivity, and heat capacity. All relevant jump conditions, including density and kinetic energy, were accounted for in the three-phase Stefan problem.

Our current work builds on our previous work to solve a more challenging problem: the three-phase Stefan problem in spherical coordinates for finite-sized particles. There are three moving interfaces in this system: the melt front, the boiling front, and the outer boundary which is in contact with the atmosphere. Although an analytical solution could not be found for this problem, we solved the governing equations using a fixed-grid sharp-interface method with second-order spatio-temporal accuracy. Using a small-time analytical solution, we predict a reasonably accurate estimate of temperature (in the three phases) and interface positions and velocities at the start of the simulation. To derive small-time analytical solutions, low- and high-density contrasts between phases are considered separately. Our numerical method is validated by reproducing the two-phase nanoparticle melting results of Font et al.~\cite{font2015nanoparticle}. Lastly, we solve the three-phase Stefan problems numerically to demonstrate the importance of kinetic energy terms during phase change of smaller (nano) particles. In contrast, these effects diminish for large particles (microns and larger).

\end{abstract}


\end{frontmatter}

\section{Introduction}\label{introduction}
The Stefan problem is concerned with the evolution of an interface between two phases while simultaneously going through a phase change, such as melting a solid metal. In 1890, Josef Stefan introduced these problems in the context of sea ice formation in a set of four papers. The foundation of Stefan's work traces back to 1831 when a similar problem was faced by Lam\'e and Clapeyron while studying the solidification of Earth's crust~\cite{vuik1993some,rubinvsteuin2000stefan}. 

Although a large body of research has been conducted to solve the Stefan problem analytically and numerically, the vast majority of studies incorporate a number of restrictive assumptions that are often justified for mathematical convenience.  These include constant latent heat, constant phase change temperature, constant thermal properties (thermal conductivity, specific heat) in each phase and an equal density in both phases~\cite{hahn2012heat, alexiades2018mathematical}. During boiling, there is a density change of 1000 or more between the liquid and vapor phases, making the assumption of equal density across phases the most unreasonable. Furthermore, most Stefan problems considered in the literature have only two phases, either solid and liquid or liquid and vapor.  

In metal manufacturing processes, such as welding and metal additive manufacturing (AM), a metal substrate or powder undergoes rapid melting, boiling, and vaporization under the action of a high-intensity laser or heat source. The standard two-phase Stefan problem cannot describe such a process. In our recent work~\cite{Mehran2025analyticalmsnbc} we introduced a three-phase Stefan problem set in half-space and Cartesian coordinates to model the simultaneous melting, solidification, boiling and condensation of a phase change material. The semi-infinite planar geometry allowed us to obtain an analytical solution to the three-phase problem that accounts for density and kinetic energy jumps during phase change. A fixed-grid sharp interface method was also presented to solve the three-phase Stefan problem with two moving interfaces in a Cartesian coordinate system. Analytical solutions were used to initialize the simulation in our prior work. The numerical method was demonstrated to be second-order accurate in both space and time.

This paper extends our previous work on the three-phase planar Stefan problem to the three-phase spherical Stefan problem while retaining appropriate density variation and kinetic energy contributions. This brings us closer to our long-term goal of developing and validating fully-resolved computational fluid dynamics (CFD) algorithms to simulate metal AM processes. In the manufacturing industry, the spherical Stefan problem has practical relevance to the production of metallic powders and spherical nanoparticles. As such, there has been extensive research in the literature regarding the spherical Stefan problem in two-phase form; see references~\cite{mccue2009micro,mccue2008classical,font2013spherically,wu2009nanoparticle,myers2020stefan_varprops}. This work contributes to the literature by introducing the three-phase version of the spherical Stefan problem and a numerical method to solve it.  

The modeled system considered here is a finite-sized spherical particle exposed to the atmosphere. Unfortunately, the finite-sized spherical geometry does not permit a similarity solution. Thus, to initialize the numerical solution for the sharp-interface method, we rely on small-time analysis of the governing equations and boundary conditions to predict the temperature in the three phases and the interface positions and velocities at the start of the simulation. Our small-time analysis of the three-phase Stefan problem can be viewed as an extension of the small-time technique introduced by Myers et al.~\cite{myers2020stefan_varprops} for modeling the two-phase nanoparticle melting. Myers et al. considered low density contrasts between the solid and liquid phases in their small-time analysis. The small-time analyses presented in this work also consider the large density contrasts between the phases. 

In Sec.~\ref{sec_3p_derivation} of our work, we mathematically formulate two-phase and three-phase spherical Stefan problems while retaining density and kinetic energy jump terms in the Stefan conditions. Governing equations and boundary conditions are non-dimensionalized and small-time analytical solutions for temperature and interface positions and velocities are derived for both two- and three-phase Stefan problems. Sec.~\ref{sec_sharp_interface_method} presents a sharp-interface numerical method using a second-order, finite difference/volume discretization, with explicit tracking of moving interfaces using an immersed-boundary (IB) method~\cite{koponen2025direct}. We then validate the proposed methodology in Sec.~\ref{sec_results} by reproducing the two-phase nanoparticle melting results of Font et al.~\cite{font2015nanoparticle}. Font el al. transform the spherical heat equations to Cartesian equations and immobilize the moving boundaries by mapping them on a fixed rectilinear domain. In contrast, we solve the governing equations directly in spherical coordinates on a fixed axisymmetric grid which allows the interfaces to freely move through the underlying mesh. Lastly, we apply the method to numerically solve the three-phase Stefan problem to quantify the influence of density and kinetic-energy terms on interface evolution. It is found that the kinetic energy terms significantly affect the melt times of smaller particles, whereas for large particles their effect diminishes significantly.

\section{Governing equations for the spherical Stefan problem }\label{sec_3p_derivation}

\begin{figure}[]
	\begin{center}
        \subfigure[Two-phase spherical Stefan problem]{\includegraphics[scale=0.18]{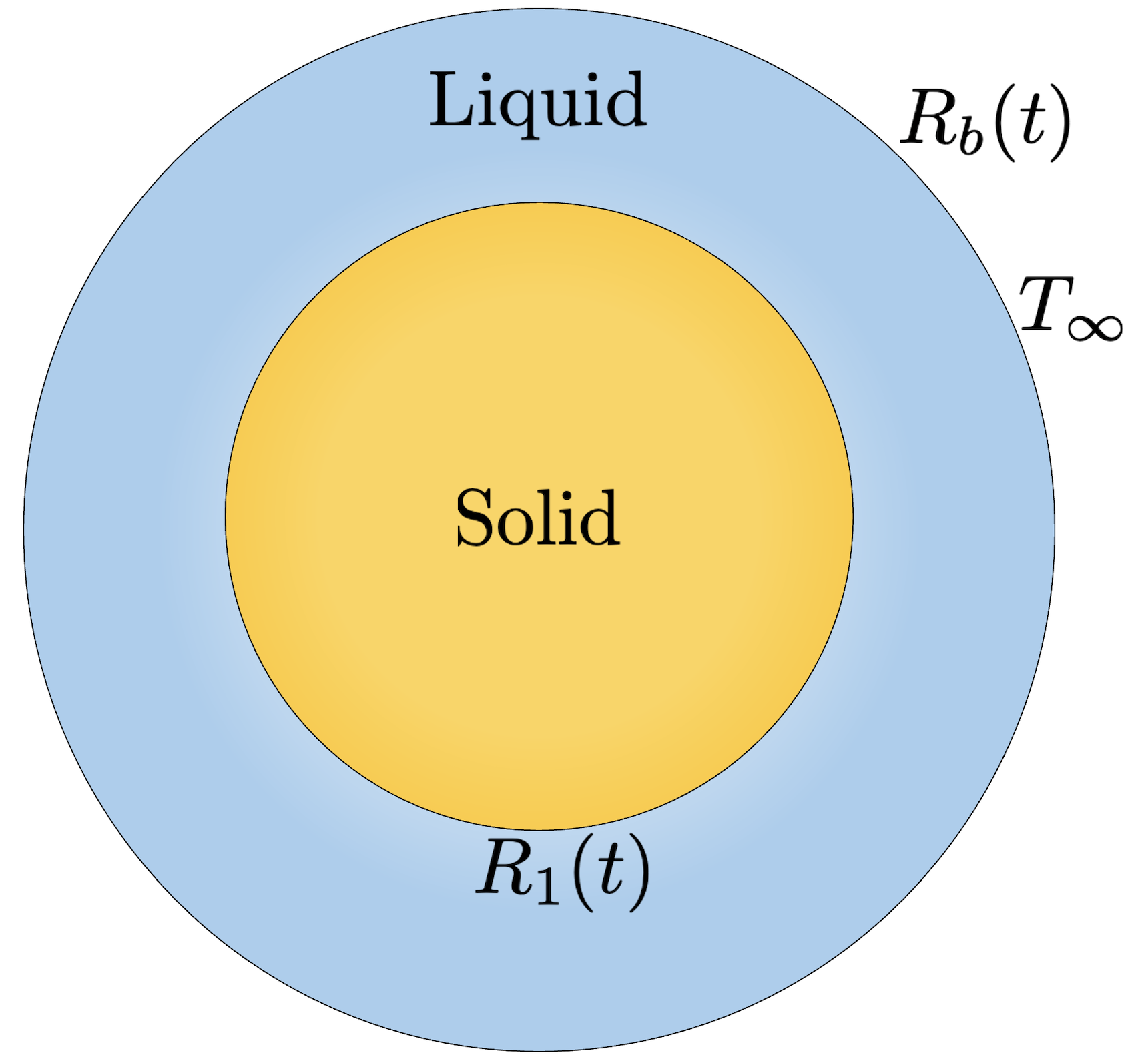}
        \label{fig_PCMGraphic_A}
        }
        \hspace*{1.7cm}
        \subfigure[Three-phase spherical Stefan problem]{\includegraphics[scale=0.18]{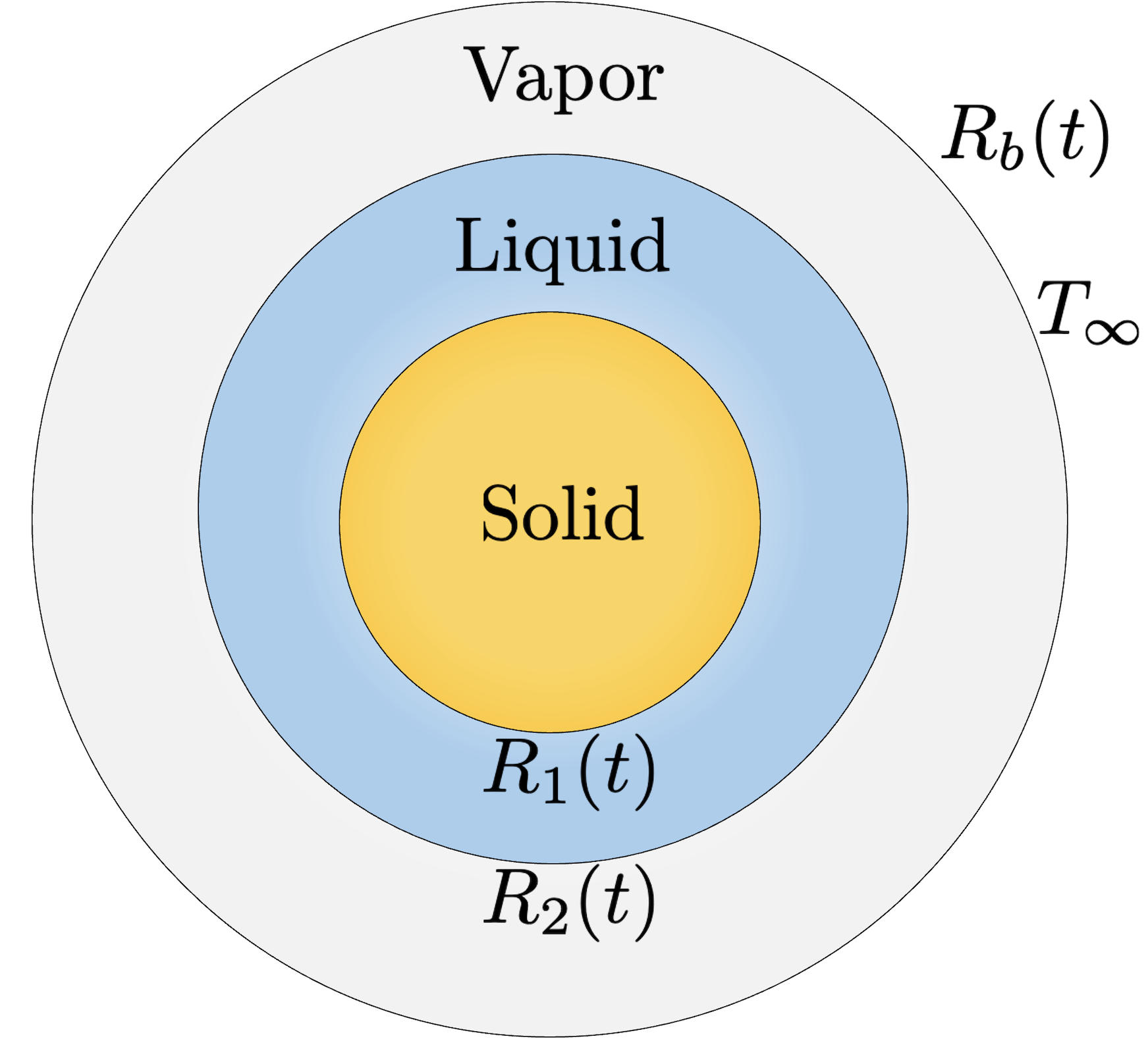}
        \label{fig_PCMGraphic_B}
        }
	\end{center}
\caption{Schematic of the \subref{fig_PCMGraphic_A} two-phase and  \subref{fig_PCMGraphic_B} three-phase spherical Stefan problem in which an initially solid PCM melts and boils, respectively due to an imposed temperature condition at the outer surface ($r = \Rb(t)$).}\label{fig_2p3pstefan_schematic}
\end{figure}

We consider a phase change material (PCM) with a melting temperature $T_m$ and a vaporization temperature $T_v$. At any time $t$, the three phases of the PCM occupy a spherically symmetric domain $\Omega := 0 \le r \le \Rb(t)$. At the initial time $t = 0$, the full domain is occupied by the solid phase. The initial radius of the solid particle is $R_b(t = 0) = \Rzero$. The outer boundary ($s_3 = \Rb(t)$) is open to the atmosphere, and it is allowed to freely expand or contract during the phase change process. Imposing a temperature of $T_\infty > T_v$ at the outer boundary melts and boils the PCM. The thermophysical properties of the solid phase are $\rhos$, $\cps$, and $\ks$, those of the liquid phase are $\rhol$, $\cpl$, and $\kl$, and those of the vapor phase are $\rhov$, $\cpv$, and $\kv$. In each phase, these properties are assumed to be constant. When the PCM melts, the melt front $s_1 = \Rone(t)$ moves inwards. The boiling front $s_2 = \Rtwo(t)$ can move inwards or outwards during the phase change process. Due to the imposed temperature $T_\infty > T_v$, the molten PCM begins to boil and forms a boiling front $ s_2 = \Rtwo(t)$; see Fig.~\ref{fig_2p3pstefan_schematic}. At $t = 0^{+}$, both fronts appear simultaneously at $r = \Rzero$. The solid-to-liquid density ratio is denoted $\rrhosl = \rhos/\rhol$, whereas its inverse is denoted $\rrhols = \rhol/\rhos$. Similarly, the liquid-to-vapor and vapor-to-liquid density ratios are denoted $\rrholv = \rhol/\rhov$ and $\rrhovl = \rhov/\rhol$, respectively. The same notation is also used to define the ratio of other thermophysical quantities, such as thermal diffusivity/conductivity.

The conservation of mass, momentum and energy equations in the solid, liquid, and vapor phases read as
\begin{subequations} \label{eqn_nstd_sp}
\begin{alignat}{2}
&\D{\rho}{t} + \V{\grad} \cdot (\rho \u ) = 0 ,\label{eqn_mass_cons} \\ 
&\D{\rho \u}{t} + \V{\grad} \cdot (\rho \u \otimes \u + p \mathbb{I}) -  \sigma \mathcal{C} \delta (\V{r} - \s) \n = 0,  \label{eqn_mom_cons} \\
&\D{}{t}\left(\rho \left[e + \frac{1}{2}| \u|^2 \right] \right) + \V{\grad} \cdot \left( \rho\left[ e + \frac{1}{2} | \u |^2 \right] \u + \q + p \u \right) - \sigma \mathcal{C} \delta (\V{r} - \s) \u \cdot \n = 0. 
 \label{eqn_energy_cons}
\end{alignat}
\end{subequations}
In the momentum Eq.~\eqref{eqn_mom_cons}, $\sigma$ is the surface energy coefficient between the two phases (vapor-liquid and liquid-solid in this context), $\mathcal{C}$ represents the mean local curvature of the interface, $\s$ represents the position of the interface, $\delta$ is the Dirac delta distribution, and $\n$ is the outward unit normal vector of the interface (pointing in the positive $r$ direction in Fig.~\ref{fig_2p3pstefan_schematic}). In the energy  Eq.~\eqref{eqn_energy_cons}, $e$ denotes the internal energy and $\q = -\kappa \V{\grad} T$ is the conductive heat flux. The melt front moves with a velocity $\uoneStar = \d{\Rone}/\d{t}$ and the boiling front moves with a velocity $\utwoStar = \d{\Rtwo}/\d{t}$. We ignore the viscous stress tensor in the momentum Eq.~\eqref{eqn_mom_cons} as shear forces are zero in a one-dimensional setting. Internal energy $e^\Gamma$ of phase $\Gamma = \{S,L,V\}$ can be expressed in terms of phasic enthalpy $h^\Gamma$ or temperature $T^\Gamma$ as
\begin{subequations} \label{eqn_enthalpies}
\begin{alignat}{2}
&\hs = \cps \left(\Ts - T_r\right) = \es + \frac{\ps}{\rhos}, \label{eqn_enthalpy_S} \\
&\hl = \cpl \left(\Tl - T_r\right) + L_m = \el + \frac{\pl}{\rhol}, \label{eqn_enthalpy_L} \\  
&\hv =  \cpv \left(\Tv - T_r\right) + L_m + L_v = \ev + \frac{\pv}{\rhov}. \label{eqn_enthalpy_V}
\end{alignat}
\end{subequations}
Here, $p^\Gamma$ is the phasic pressure, $T_r$ is the reference temperature at which enthalpies are measured, and $L_m$ and $L_v$ are the latent heats of melting and vaporization, respectively.

\subsection{Two phase spherical Stefan problem} \label{sec_2p_stefan_eqs}
We consider a simplified scenario where the PCM melts only during phase change. This is achieved by setting $T_\infty$ greater than $T_m$ but less than $T_v$. A similar two-phase spherical Stefan problem has been considered in the literature~\cite{font2015nanoparticle}. The two-phase problem serves to (i) introduce the non-dimensional parameters of the problem; (ii) introduce a small-time analytical solution method; and (iii) validate our fixed-grid sharp-interface numerical method with a coordinate transformed numerical solution reported in the literature. 

To find the evolution of temperature in the solid ($\Ts(r,t)$) and liquid ($\Tl(r,t)$) phases as a result of melting, we need to solve the temperature
equation in the two domains. This is obtained by expressing the internal energy $e$ in terms of temperature (see Eqs.~\eqref{eqn_enthalpy_S} and~\eqref{eqn_enthalpy_L}), and
neglecting terms related to pressure work and viscous dissipation in Eq.~\eqref{eqn_energy_cons}. The resulting temperature equation in the solid and liquid phases reads as 
\begin{subequations}  \label{eq:2P-dim-heat}
  \begin{align}
    \rhos \cps\left(\D{\Ts}{t} + \us\,\D{\Ts}{r}\right)&= \frac{\ks}{r^2}\,\D{}{r}\left(r^2 \D{\Ts}{r}\right), && 0 \le r<\Rone(t) \label{eq:2P-dim-heat-S}\\ 
    \rhol \cpl\left(\D{\Tl}{t} + \ul\,\D{\Tl}{r}\right)&=  \frac{\kl}{r^2}\,\D{}{r}\left(r^2 \D{\Tl}{r}\right). && \Rone(t)<r \le \Rb(t) \label{eq:2P-dim-heat-L}
  \end{align}
\end{subequations}
Five boundary conditions are needed to solve Eqs.~\eqref{eq:2P-dim-heat} to determine $\Ts(r,t)$, $\Tl(r,t)$, and $\Rone(t)$. 

The temperature $\Tim$ at the S-L interface differs from the bulk melting temperature $T_m$ due to the curvature and surface energy/tension effects. The two temperatures are related by the Gibbs-Thomson relation
\begin{equation}
  \Tim(\Rone(t)) = T_m - \frac{\sigma_{\rm SL} T_m}{\rhos L_m}\,\frac{2}{\Rone(t)}. \label{eq:GT-melting-dim}
\end{equation}
Here, $\sigma_{\rm SL}$ denotes the surface energy coefficient between the solid and liquid phases. At any time $t$ during the melting process, the temperature in the liquid and solid phase near the melt front is the same: $\Ts(\Rone(t),t) = \Tl(\Rone(t),t) = \Tim(\Rone(t))$. The outer boundary is maintained at an ambient temperature of $\Tl(\Rb(t),t) = \Tinf$. Solid phase temperature at the center of the sphere satisfies the homogeneous Neumann boundary condition $\left.\D{\Ts}{r}\right|_{r=0} = 0$. The Stefan condition (and the fifth boundary condition) for the melt front is obtained by applying the Rankine-Hugoniot jump condition to the energy equation across the solid-liquid interface
\begin{equation} 
 \ks \D{\Ts}{r}\Bigg|_{\Rone}- \kl \D{\Tl}{r}\Bigg|_{\Rone} = \rhos \uoneStar \bigg[(\cpl - \cps)(\Tim - \Tr) + L_m - \frac{1}{2}\bigg(1 - (\rrhosl)^2\bigg)({\uoneStar})^2\bigg]. \label{eq:Stefan-2P-dim}
\end{equation}
We refer the reader to our prior work~\cite{Mehran2025analyticalmsnbc} for the derivation of the Stefan condition(s) for the two- and three-phase Stefan problems.  

Radii $\Rone(t)$ and $R_b(t)$ are related through mass conservation of the PCM:
\begin{align}
&\frac{4}{3}\pi \rhos \Rzero^3 = \frac{4}{3}\pi \rhos \Rone^3(t) + \frac{4}{3}\pi \rhol (R_b^3(t) - \Rone^3(t))  \nonumber \\
\hookrightarrow\, & R_b^3(t) = \rrhosl \Rzero^3 + (1 - \rrhosl)\Rone^3(t). \label{eq:2PRbR1-dim}
\end{align}

The velocity in the solid phase is zero, i.e., $\us(\Omegas,t) \equiv 0$, whereas in the liquid phase is a function of time and position. $\ul$ is obtained by integrating the mass balance Eq.~\eqref{eqn_mass_cons} in the liquid domain. This yields
\begin{subequations}
\begin{align}
&\D{}{r}\left(r^2 \ul \right) = 0,  \\
\hookrightarrow\, & \ul(r,t) = \frac{A(t)}{r^2},
\end{align}
\end{subequations}
in which $A$ is a yet to be determined function of time. Applying the Rankine-Hugoniot jump condition across the solid-liquid interface for the mass balance equation relates $A(t)$ to the interface position $\Rone(t)$ and velocity $\uoneStar(t)$: 
\begin{subequations}
\begin{align}
&(\rhol - \rhos)\dd{\Rone}{t} = \rhol \ul\bigg|_\Rone, \\
\hookrightarrow\, &  \ul\bigg|_\Rone = \frac{A}{\Rone^2} = \left( 1 - \rrhosl\right)\dd{\Rone}{t}, \\
\hookrightarrow\, & \ul(r,t) = \frac{\Rone^2\left( 1 - \rrhosl\right)}{r^2}\uoneStar. \label{eq:ul(r,t)}
\end{align}
\end{subequations}

In order to determine the relevant parameters of the problem, we non-dimensionalize the governing equations and the boundary conditions. Non-dimensional variables are indicated with a hat $\widehat{(.)}$ symbol. We define 
\begin{equation}
t = \tau \that, \quad r = \Rzero\,\rhat, \quad T = T_r + \DeltaT\,\That, 
\end{equation}
in which the time, length, and temperature scales are $\tau$, $\Rzero$, and $\Delta T$, respectively. For externally imposed temperature condition, the temperature scale can be defined as $\Delta T = \Tinf - T_r$. The length and time scales determine the velocity scale of the problem, i.e., $u = \frac{\Rzero}{\tau} \uhat$. The time scale $\tau = \Rzero^2/\alphal$ is chosen based on (thermal) diffusion in the liquid phase; this is not a unique choice. Here, $\alphal = \kl/\rhol\cpl$ is the thermal diffusivity of the liquid phase. The length scale $\Rzero$ is taken to be the initial radius of the particle. By using the above scales, the non-dimensional temperature equation in the liquid and solid phases reads as
\begin{subequations} \label{eq:2P-nondim-heat-all}
  \begin{align}
    \D{\Thats}{\that} + \uhats\,\D{\Thats}{\rhat}&= \alpha^{\rm SL}\,\frac{1}{\rhat^2}\,\D{}{\rhat}\left(\rhat^2 \D{\Thats}{\rhat}\right), && 0 \le \rhat<\Rhatone(\that) \label{eq:2P-nondim-S}\\
    \D{\Thatl}{\that} + \uhatl\,\D{\Thatl}{\rhat}&= \frac{1}{\rhat^2}\,\D{}{\rhat}\left(\rhat^2 \D{\Thatl}{\rhat}\right). && \Rhatone(\that)<\rhat \le \Rhatb(\that)     \label{eq:2P-nondim-L}
  \end{align}
\end{subequations}
Here, $\alpha^{\rm SL} = \alphas / \alphal$ denotes solid-to-liquid thermal diffusivity ratio,  $\Rhatone = \Rone/\Rzero$, and $\Rhatb = \Rb/\Rzero$. 

The boundary condition equations are also non-dimensionalized using the same scales. To wit, the non-dimensional version of the interface temperature Eq.~\eqref{eq:GT-melting-dim} reads as 
\begin{equation}
  \Thatim = \Thatm - \frac{\Gammam}{\Rhatone}, \label{eq:GT-melting-ndim}
\end{equation}
in which $\Thatim  = (\Tim - T_r)/\Delta T$, $\Thatm  = (T_m - T_r)/\Delta T$, and $\Gammam = (2\sigma_{\rm SL} T_m)/(\rhos L_m \Rzero\Delta T)$. Proceeding analogously, the non-dimensional version of the Stefan condition Eq.~\eqref{eq:Stefan-2P-dim} reads as
\begin{equation}
  \label{eq:Stefan-2P-ndim}
  \kappa^{\rm SL}\left.\D{\Thats}{\rhat}\right|_{\Rhatone}- \left.\D{\Thatl}{\rhat}\right|_{\Rhatone}= \rrhosl \betam\left[1 + \gammam \Thatim - \frac{1}{2}\bigl( 1 - (\rrhosl)^2 \bigr)\deltam \uhatone^2\right]\uhatone.
\end{equation}
Here, $\kappa^{\rm SL} = \ks / \kl$ denotes solid-to-liquid thermal conductivity ratio, $ \betam = L_m/(\cpl\,\Delta T)$ is the (liquid) Stefan number, $\gammam = (\cpl - \cps)\Delta T/L_m$ is the inverse Stefan number, and $\deltam = \Rzero^2/(L_m \tau^2)$ is a measure of kinetic energy of the melt front relative to the latent heat of melting.

\subsection{Small-time analytical solution to the two-phase spherical Stefan problem} \label{sec_2P_small_time}

In order to initialize the numerical simulation we need to understand the behavior of the system (interface velocity, phase temperature, etc.) at the initial stages of the melting process. At the initial times of the phase change process, it is reasonable to expect that the melt front will stay close to the outer boundary, i.e., $\Rhatone \approx 1$, and that the initial thickness of the liquid layer will also be small. Based on these assumptions, we re-scale the spatial and temporal coordinates. Specifically, utilizing the smallness parameter $0<\varepsilon\ll 1$, we define small-time position and time coordinates indicated with a tilde $\widetilde{(.)}$ symbol: 
\begin{equation}
\rhat = 1 - \varepsilon \rtil, \quad \Rhatone = 1 - \varepsilon \Rtilone, \quad \Rhatb = 1 -\varepsilon\Rtilb, \quad \that = \varepsilon \ttil. 
\end{equation}
Furthermore, the small-time velocity of the interface is assumed to be constant, $\d \Rtilone / \d \ttil = P$. Thus, the dimensionless position and velocity of the interface at the initial times can be written as 
\begin{subequations} \label{eq:2p-smalltime-pos-vel}
\begin{align}
   &\Rhatone(\that)  = 1 - \varepsilon P\ttil = 1 - P\that,  \\
   &\uhatone  = \dd{\Rhatone}{\that}  = -P.
\end{align}
\end{subequations}
$\rtil$, $\Rtilone$, $\ttil$, and $P$ are assumed to be $\mathcal{O}(1)$. $\Rtilone$ and $\Rtilb$ are related to each other through Eq.~\eqref{eq:2PRbR1-dim} as follows:
\begin{align}
\Rhatb^3 &= \rrhosl + (1 - \rrhosl)\Rhatone^3,  \nonumber \\
\Rhatb^3 &= \rrhosl + (1 - \rrhosl)(1 - \varepsilon\Rtilone)^3, \nonumber \\
\Rhatb^3  &\approx \rrhosl + (1 - \rrhosl)(1 - 3\varepsilon\Rtilone), \nonumber \\
\Rhatb^3 &= 1 - 3\varepsilon(1-\rrhosl)\Rtilone, \nonumber \\
\Rhatb   &= \bigg[1 - 3\varepsilon(1-\rrhosl)\Rtilone\bigg]^{\frac{1}{3}}, \nonumber \\
1 -\varepsilon\Rtilb &\approx 1 - \varepsilon(1-\rrhosl)\Rtilone, \nonumber \\
\implies \Rtilb &= (1-\rrhosl)\Rtilone. \label{eq:2P-Rbtil}
\end{align}

Next we apply these small-time scalings to the dimensionless temperature Eq.~\eqref{eq:2P-nondim-L} for the liquid phase. Various terms in the temperature equation scale as
\begin{subequations} \label{eq:small-time-scaling}
  \begin{align}
    &\frac{\partial \Thatl}{\partial \that}= \frac{1}{\varepsilon}\,\frac{\partial \Ttill}{\partial \ttil},\\
    &\frac{\partial \Thatl}{\partial \rhat}= -\frac{1}{\varepsilon}\,\frac{\partial \Ttill}{\partial \rtil}, \\
    &\uhatl = \frac{(1-\rrhosl)\Rhatone^2}{\rhat^2}\uhatone = \frac{(1-\rrhosl)(1-\varepsilon\Rtilone)^2}{(1-\varepsilon\rtil)^2}(-P),  \label{eq:small-time-ul-scaling}\\
    &\rhat^2 = (1 - \varepsilon\rtil)^2 = 1 - 2\varepsilon\rtil + \mathcal{O}(\varepsilon^2), \\
    & \rhat^2\frac{\partial \Thatl}{\partial \rhat} = (1 - 2\varepsilon\rtil + \cdots) \left(-\frac{1}{\varepsilon}\, \frac{\partial \Ttill}{\partial \rtil}\right)= -\frac{1}{\varepsilon}\,\frac{\partial\Ttill}{\partial \rtil}+ \mathcal{O}(1),\\
    &\frac{\partial}{\partial \rhat}\left(\rhat^2\frac{\partial \Thatl}{\partial \rhat}\right)= -\frac{1}{\varepsilon}\,\frac{\partial}{\partial \rtil}\left(-\frac{1}{\varepsilon}\,\frac{\partial\Ttill}{\partial \rtil}+ \mathcal{O}(1)\right)= \frac{1}{\varepsilon^2}\,\frac{\partial^2 \Ttill}{\partial \rtil^2}+ \mathcal{O}\left(\frac{1}{\varepsilon}\right),\\
    &\frac{1}{\rhat^2}\frac{\partial}{\partial \rhat}\left(\rhat^2\frac{\partial \Thatl}{\partial \rhat}\right)= \left[1 + \mathcal{O}(\varepsilon)\right]\left[\frac{1}{\varepsilon^2}\,\frac{\partial^2 \Ttill}{\partial \rtil^2} + \mathcal{O}\left(\frac{1}{\varepsilon}\right)\right]= \frac{1}{\varepsilon^2}\, \frac{\partial^2 \Ttill}{\partial \rtil^2}+ \mathcal{O}\left(\frac{1}{\varepsilon}\right).
  \end{align}
\end{subequations}

\noindent According to the magnitude of the solid-liquid density ratio $\rrhosl \approx 1$ or $\rrhosl \gg 1$, the liquid domain velocity $\uhatl$ (Eq.~\eqref{eq:small-time-ul-scaling}) scales as follows:
\begin{subequations} \label{eq:2P-ul-scaling}
  \begin{align}
\uhatl &\approx  (1-\rrhosl)(-P) \approx P = \mathcal{O}(1), & \text{[low density ratio scaling]} \label{eq:2P-ul-ldr-scaling} \\
\uhatl &\approx  \left(1-\frac{\rrhosltil}{\varepsilon} \right)(-P) \approx \frac{ \rrhosltil} {\varepsilon} P = \mathcal{O}\left(\frac{1}{\varepsilon}\right), & \text{[high density ratio scaling]} \label{eq:2P-ul-hdr-scaling}
\end{align}
\end{subequations}
in which the large solid-liquid density ratio is scaled as $\rrhosl = \rrhosltil/\varepsilon$ with $\rrhosltil \sim \mathcal{O}(1)$. Therefore, we obtain two different small-time solutions for low and high density ratios $\rrhosl$.

\subsubsection{Small-time low-density-ratio solution (LDRS) to the two-phase spherical Stefan problem} \label{sec_2p-ldrs}

Substituting Eqs.~\eqref{eq:small-time-scaling} and Eq.~\eqref{eq:2P-ul-ldr-scaling} into Eq.~\eqref{eq:2P-nondim-L} 
\begin{equation}
  \frac{1}{\varepsilon}\,\frac{\partial \Ttill}{\partial \ttil} - \frac{P}{\varepsilon}\,\frac{\partial \Ttill}{\partial \rtil}= \frac{1}{\varepsilon^2}\,\frac{\partial^2 \Ttill}{\partial \rtil^2}+ \mathcal{O}\left(\frac{1}{\varepsilon}\right),
\end{equation}
yields the leading-order term that determines the small-time temperature in the liquid phase: 
\begin{equation}
  \label{eq:2p-smalltime-ld-TlPDE}
  \frac{\partial^2 \Ttill}{\partial \rtil^2} = 0.
\end{equation}
The general solution of Eq.~\eqref{eq:2p-smalltime-ld-TlPDE}  is of the form: 
\begin{equation}
  \label{eq:2p-smalltime-linear}
  \Ttill(\rtil,\ttil)= C_1(\ttil)\,\rtil + C_2(\ttil).
\end{equation}
The boundary condition at the outer surface enclosing the liquid phase is $\Ttill(\Rtilb,\ttil) = \Thatinf$, in which $\Thatinf = (\Tinf-\Tr)/\Delta T$. The interface temperature condition requires $\Ttill(\Rtilone,\ttil) = \Ttilim$. The leading-order interface temperature is obtained from Eq.~\eqref{eq:GT-melting-ndim} as $\Ttilim = \Thatm - \Gammam$.  Imposing the two temperature conditions to Eq.~\eqref{eq:2p-smalltime-linear} gives
\begin{subequations}
  \label{eq:2p-TL-smalltime-A1B1}
  \begin{align}
      C_1\Rtilb + C_2 &= C_1(1-\rrhosl)\Rtilone + C_2 = \Thatinf, \\
      C_1\Rtilone + C_2 &= \Ttilim.
  \end{align} 
\end{subequations}
Solving for $C_1$ and $C_2$ from the equations above yields small-time temperature in the liquid phase as
\begin{subequations}
\begin{align}
  &\Ttill(\rtil,\ttil)= \frac{\Thatinf - \Ttilim}{\rrhosl}\left(1- \frac{\rtil}{\Rtilone}\right) + \Ttilim, \label{eq:2pTl-ld-smalltime-coords} \\
  \hookrightarrow\, &\Thatl(\rhat,\that) = \frac{\Thatinf - \Ttilim}{\rrhosl}\left(1- \frac{1-\rhat}{1-\Rhatone}\right) + \Ttilim. \label{eq:2pTl-ld-nondim-coords}
\end{align}
\end{subequations}
Note that Eq.~\eqref{eq:2pTl-ld-nondim-coords} is obtained from Eq.~\eqref{eq:2pTl-ld-smalltime-coords} by substituting $\rtil = (1 - \rhat)/\varepsilon$ and $\Rtilone = (1 - \Rhatone)/\varepsilon$. 

Analogously, the leading-order term in the solid phase's temperature equation is 
\begin{equation}
  \label{eq:2p-smalltime-TsPDE}
  \frac{\partial^2 \Ttils}{\partial \rtil^2} = 0.
\end{equation}
Since $\D{\Thats}{\rhat}\bigg|_{\rhat = 0} = 0$, and $\Thats(\Rhatone,\that) = \Thatim$, at initial times the solid phase has a uniform temperature throughout its domain per Eq.~\eqref{eq:2p-smalltime-TsPDE}: 
\begin{equation}
\Ttils(\rtil, \ttil) = \Thats(\rhat, \that) = \Ttilim. \label{eq:2P-smalltime-Ts}
\end{equation}

In order to determine the small-time position and velocity of the interface using Eqs.~\eqref{eq:2p-smalltime-pos-vel}, the value of $P$ needs to be determined. This is obtained by substituting the small-time solutions $\Thatl$ and $\Thats$ into the Stefan Eq.~\eqref{eq:Stefan-2P-ndim}
\begin{align}
 & \kappa^{\rm SL}\left.\D{\Thats}{\rhat}\right|_{\Rhatone}- \left.\D{\Thatl}{\rhat}\right|_{\Rhatone}= \rrhosl \betam\left[1 + \gammam \Thatim - \frac{1}{2}\bigl( 1 - (\rrhosl)^2 \bigr)\deltam \uhatone^2\right]\uhatone, \nonumber \\
 \hookrightarrow\, & \frac{\Thatinf - \Ttilim}{\rrhosl}\left(\frac{1}{P\that}\right) = \rrhosl \betam\left[1 + \gammam \Ttilim - \frac{1}{2}\bigl( 1 - (\rrhosl)^2 \bigr)\deltam P^2\right]P. \label{eq:2P-ld-Psol}
\end{align}
Eq.~\eqref{eq:2P-ld-Psol} is a quartic equation in $P$ that needs to solved numerically using a root finding algorithm. In our numerical experiments we take a small value of the initial nondimensional time $\that_{\rm init} \le 10^{-3}$.  Once $P$ is determined by solving Eq.~\eqref{eq:2P-ld-Psol}, the small-time low-density-ratio solution (LDRS) to the two-phase Stefan problem is complete. 

\subsubsection{Small-time high-density-ratio solution (HDRS) to the two-phase spherical Stefan problem} \label{sec_2p-hdrs}

Substituting Eqs.~\eqref{eq:small-time-scaling} and Eq.~\eqref{eq:2P-ul-hdr-scaling} into Eq.~\eqref{eq:2P-nondim-L} 
\begin{equation}
  \frac{1}{\varepsilon}\,\frac{\partial \Ttill}{\partial \ttil} - \frac{\rrhosltil}{\varepsilon^2}P\,\frac{\partial \Ttill}{\partial \rtil}= \frac{1}{\varepsilon^2}\,\frac{\partial^2 \Ttill}{\partial \rtil^2}+ \mathcal{O}\left(\frac{1}{\varepsilon}\right),
\end{equation}
yields the leading-order term that determines the small-time temperature in the liquid phase: 
\begin{equation}
  \label{eq:2p-smalltime-hd-TlPDE}
  \frac{\partial^2 \Ttill}{\partial \rtil^2} = G\frac{\partial \Ttill}{\partial \rtil}.
\end{equation}
Here, $G = -P \, \rrhosltil$. Integrating Eq.~\eqref{eq:2p-smalltime-hd-TlPDE} twice with respect to $\rtil$ gives the general solution of the form
\begin{subequations}
\begin{align}
  \Ttill(\rtil,\ttil) &= C_3(\ttil)\,\frac{\exp(G\rtil)}{G} + C^'_4(\ttil), \\
  &= C_4 + C_3 \bigg[ \rtil + \frac{G\rtil^2}{2} + \cdots \bigg], 
\end{align}
\end{subequations}
in which we have retained up to quadratic term of the exponential function. $C_3$ and $C_4$ functions are determined by imposing the boundary conditions: $\Ttill(\Rtilb,\ttil) = \Thatinf$ and $\Ttill(\Rtilone,\ttil) = \Ttilim$. This yields
\begin{subequations}
\begin{align}
&\Ttill(\rtil,\ttil) = \Thatinf - (\Thatinf - \Ttilim)\frac{(\Rtilb - \rtil) + \frac{G}{2}(\Rtilb^2 - \rtil^2)}{(\Rtilb - \Rtilone) + \frac{G}{2}(\Rtilb^2 - \Rtilone^2)  }, \label{eq:2pTl-hd-smalltime-coords}  \\
\hookrightarrow\, &\Thatl(\rhat,\that) = \Thatinf - (\Thatinf - \Ttilim)\frac{(1-\rrhosl)^2(1-\Rhatone)^2 - (1-\rhat)^2}{(1-\rrhosl)^2(1-\Rhatone)^2 - (1-\Rhatone)^2}. \label{eq:2pTl-hd-nondim-coords}
\end{align}
\end{subequations}
Eq.~\eqref{eq:2pTl-hd-nondim-coords} is obtained from Eq.~\eqref{eq:2pTl-hd-smalltime-coords} by substituting $\rtil = (1 - \rhat)/\varepsilon$ and $\Rtilone = (1 - \Rhatone)/\varepsilon$, and taking the limit $\varepsilon \rightarrow 0$. The temperature in the solid domain remains the same as written in Eq.~\eqref{eq:2P-smalltime-Ts}.

Finally, substituting the small-time HDRS $\Thatl$ and $\Thats$ into the Stefan Eq.~\eqref{eq:Stefan-2P-ndim} gives us the equation for calculating $P$ for the high density ratio case:
\begin{equation}
\frac{\Thatinf - \Ttilim}{\big[1-(1-\rrhosl)^2\big]}\left(\frac{2}{P\that}\right) = \rrhosl \betam\left[1 + \gammam \Ttilim - \frac{1}{2}\bigl( 1 - (\rrhosl)^2 \bigr)\deltam P^2\right]P. \label{eq:2P-hd-Psol}
\end{equation}

\subsection{Three phase spherical Stefan problem}
For the the three-phase Stefan problem  there are three interfaces present in the domain: solid-liquid interface $\Rone(t)$, liquid-vapor interface $\Rtwo(t)$, and outer free boundary $\Rb(t)$ with the condition $0 < \Rone(t) < \Rtwo(t) < \Rb(t)$. The dimensional form of the temperature equation in the solid, liquid, and vapor phases reads as 
\begin{subequations}  \label{eq:3P-dim-heat}
  \begin{align}
    \rhos \cps\left(\D{\Ts}{t} + \us\,\D{\Ts}{r}\right)&= \frac{\ks}{r^2}\,\D{}{r}\left(r^2 \D{\Ts}{r}\right),&& 0 \le r<\Rone(t) \label{eq:3P-dim-heat-S}\\
       \rhol \cpl\left(\D{\Tl}{t} + \ul\,\D{\Tl}{r}\right)&=  \frac{\kl}{r^2}\,\D{}{r}\left(r^2 \D{\Tl}{r}\right), && \Rone(t)<r<\Rtwo(t) \label{eq:3P-dim-heat-L} \\
       \rhov \cpv\left(\D{\Tv}{t} + \uv\,\D{\Tv}{r}\right)&=  \frac{\kv}{r^2}\,\D{}{r}\left(r^2 \D{\Tv}{r}\right). && \Rtwo(t)<r \le\Rb(t) \label{eq:3P-dim-heat-V}
  \end{align}
\end{subequations}
Using the same scales for length, time, and temperature as defined in Sec.~\ref{sec_2p_stefan_eqs}, the dimensionless form of Eqs.~\eqref{eq:3P-dim-heat} reads as
\begin{subequations}\label{eq:3P-non-dim-heat}
    \begin{align}
        \D{\Thats}{\that} + \uhats\,\D{\Thats}{\rhat}&= \alpha^{\rm SL}\,\frac{1}{\rhat^2}\,\D{}{\rhat}\left(\rhat^2 \D{\Thats}{\rhat}\right), && 0\le \rhat<\Rhatone(\that) \label{eq:3P-nondim-S}\\
    \D{\Thatl}{\that} + \uhatl\,\D{\Thatl}{\rhat}&= \frac{1}{\rhat^2}\,\D{}{\rhat}\left(\rhat^2 \D{\Thatl}{\rhat}\right), && \Rhatone(\that)<\rhat<\Rhattwo(\that)     \label{eq:3P-nondim-L}\\
    \D{\Thatv}{\that} + \uhatv\,\D{\Thatv}{\rhat}&= \alpha^{\rm VL}\,\frac{1}{\rhat^2}\,\D{}{\rhat}\left(\rhat^2 \D{\Thatv}{\rhat}\right). && \Rhattwo(\that)<\rhat \le \Rhatb(\that)     \label{eq:3P-nondim-V}
    \end{align}
\end{subequations}
The dimensionless form of the Gibbs-Thomson relation that defines temperature at the boiling front reads as
\begin{equation}
  \Thativ = \That_v - \frac{\Gammav}{\Rhattwo}, \label{eq:GT-vap-ndim}
\end{equation}
in which $\Thativ  = (\Tiv - T_r)/\Delta T$, $\That_v  = (T_v - T_r)/\Delta T$, and $\Gammav = (2\sigma_{\rm LV} T_v)/(\rhol L_v \Rzero\Delta T)$. The dimensionless form of the solid-liquid interface temperature $\Thatim$ remains the same as written in Eq.~\eqref{eq:GT-melting-ndim}. The temperature at the outer boundary is maintained at $\Tinf$, so that the condition $\Thatv(\Rhatb,\that) = \Thatinf$ holds at all times. By applying the Rankine-Hugoniot jump condition to the energy equation, we obtain the Stefan condition for the boiling front:
\begin{alignat}{2}\label{eq:Stefan-3P-dim}
&\kl \D{\Tl}{r}\Bigg|_{\Rtwo(t)}- \kv \D{\Tv}{r}\Bigg|_{\Rtwo(t)} = \bigg(\rhol \utwoStar - (\rhol-\rhos)\uoneStar \bigg) \Bigg[(\cpv-\cpl)(\Tiv-\Tr) + L_v -\frac{1}{2}\Big\{(1-\rrhosl)\uoneStar\Big\}^2 + \nonumber \\ &\quad\quad  +\frac{1}{2}\Big\{(\rrholv-\rrhosv)\uoneStar+(1-\rrholv)\utwoStar\Big\}^2-\Big\{(\rrholv-\rrhosv)\uoneStar+(1-\rrholv)\utwoStar-(1-\rrhosl)\uoneStar\Big\}\utwoStar\Bigg].
\end{alignat}
Our prior work~\cite{Mehran2025analyticalmsnbc} provides a detailed derivation of the Stefan conditions for the three-phase case. Eq.~\eqref{eq:Stefan-3P-dim} when non-dimensionalized with the scaling parameters defined previously gives
\begin{equation}\label{eq:Stefan-3P-ndim}
\left.\D{\Thatl}{\rhat}\right|_{\Rhattwo} - \kappa^{\rm VL}\left.\D{\Thatv}{\rhat}\right|_{\Rhattwo} = \betav\Big(\uhattwo-(1-\rrhosl)\uhatone\Big)\bv,
\end{equation}
in which $\bv$ is defined as
\begin{alignat}{2}\label{eq:bv-def}
    \bv &= \Bigg[1+\gammav \Thativ-\frac{1}{2}(1-\rrhosl)^2\deltav\,\uhatone^2+\frac{1}{2}\deltav\Big\{(\rrholv-\rrhosv)\uhatone+(1-\rrholv)\uhattwo\Big\}^2 - \nonumber \\ 
    &\quad\quad \deltav\Big\{(\rrholv-\rrhosv)\uhatone+(1-\rrholv)\uhattwo-(1-\rrhosl)\uhatone\Big\}\uhattwo \Bigg].
\end{alignat}
Here, $\kappa^{\rm VL} = \kv / \kl$ denotes vapor-to-liquid thermal conductivity ratio, $ \betav = L_v/(\cpl\,\Delta T)$ is the (vapor) Stefan number, $\gammav = (\cpv - \cpl)\Delta T/L_v$ is the inverse Stefan number, and $\deltav = \Rzero^2/(L_v \tau^2)$ is a measure of kinetic energy of the boiling front relative to the latent heat of vaporization. The dimensional and dimensionless version of the Stefan equation for the melt front remains the same as written in Eqs.~\eqref{eq:Stefan-2P-dim} and~\eqref{eq:Stefan-2P-ndim}, respectively. 

The velocity in the solid phase is zero, i.e., $\us(\Omegas,t) \equiv 0$, whereas in the liquid phase remains the same as written in Eq.~\eqref{eq:ul(r,t)}. The velocity in the vapor phase $\uv$ is obtained by integrating the mass balance Eq.~\eqref{eqn_mass_cons} in the vapor domain. This yields
\begin{subequations}
\begin{align}
&\D{}{r}\left(r^2 \uv \right) = 0,  \\
\hookrightarrow\, & \uv(r,t) = \frac{B(t)}{r^2},
\end{align}
\end{subequations}
in which $B$ is a yet to be determined function of time. Applying the Rankine-Hugoniot jump condition to the mass balance equation across the liquid-vapor interface relates $B(t)$ to the interfacial positions $\Rone(t)$ and $\Rtwo(t)$, and interfacial velocities $\uoneStar(t)$ and $\utwoStar(t)$: 
\begin{subequations}
\begin{align}
&(\rhol - \rhov)\dd{\Rtwo}{t} = (\rhol \ul - \rhov \uv)\bigg|_\Rtwo, \\
\hookrightarrow\, &  (\rhol - \rhov)\dd{\Rtwo}{t} =  \rhol(1-\rrhosl)\frac{\Rone^2}{\Rtwo^2}\dd{\Rone}{t} - \rhov\frac{B}{\Rtwo^2}, \\
\hookrightarrow\, & B(t) = \bigg[ (\rrholv - \rrhosv)\Rone^2 \dd{\Rone}{t} - (\rrholv - 1)\Rtwo^2\dd{\Rtwo}{t} \bigg], \\
\implies & \uv(r,t) = \bigg[ (\rrholv - \rrhosv)\frac{\Rone^2}{r^2} \uoneStar - (\rrholv - 1)\frac{\Rtwo^2}{r^2}\utwoStar \bigg], \label{eq:uv(r,t)} \\
\hookrightarrow\, & \uhatv(\rhat,\that) =  \bigg[ (\rrholv - \rrhosv)\frac{\Rhatone^2}{\rhat^2} \uhatone - (\rrholv - 1)\frac{\Rhattwo^2}{\rhat^2}\uhattwo \bigg].  \label{eq:uhatv(r,t)}
\end{align}
\end{subequations}

Radii $\Rb(t)$, $\Rone(t)$, and $\Rtwo(t)$ are related to each other through the mass conservation of the PCM (similar to the two-phase case). Alternatively, Eq.~\eqref{eq:uv(r,t)} can be used to derive the relationship between the three radii:
\begin{subequations}
\begin{align}
& \dd{\Rb}{t} = \uv(\Rb,t) = \bigg[ (\rrholv - \rrhosv)\frac{\Rone^2}{\Rb^2} \dd{\Rone}{t} - (\rrholv - 1)\frac{\Rtwo^2}{\Rb^2}\dd{\Rtwo}{t} \bigg], \\
\hookrightarrow\, &   \Rb^2\,\d\Rb = (\rrholv - \rrhosv) \Rone^2\, \d\Rone - (\rrholv - 1)\Rtwo^2\, \d\Rtwo, \label{eq:3P-Rb-indefin} \\
\implies\, & \Rb^3 = a(\Rone^3 - \Rzero^3) - b(\Rtwo^3 - \Rzero^3) + \Rtwo^3, \label{eq:3P-Rb-defin} \\
\hookrightarrow\, & \Rhatb^3 = a(\Rhatone^3 - 1) - b(\Rhattwo^3 - 1) + \Rhattwo^3, \label{eq:3P-Rb-ndim}
\end{align}
\end{subequations}
in which we have used the initial condition $\Rone = \Rtwo = \Rb = \Rzero$ at $t = 0$ while integrating Eq.~\eqref{eq:3P-Rb-indefin}, and defined $a = (\rrholv - \rrhosv)$ and $b = \rrholv$ in Eqs.~\eqref{eq:3P-Rb-defin} and~\eqref{eq:3P-Rb-ndim}. 

\subsection{Small-time analytical solution to the three-phase spherical Stefan problem} \label{sec_3P_small_time}

Having discussed the governing equations and the boundary conditions for the three-phase Stefan problem, we derive the small-time analytical solution to initialize the simulation. At the initial times we assume $\Rhatone \approx \Rhattwo \approx \Rhatb \approx 1$. By using this ansatz, same as the one used for the two-phase case, we define small-time position and time coordinates using $0<\varepsilon\ll 1$: 
\begin{equation}
\rhat = 1 - \varepsilon \rtil, \quad \Rhatone = 1 - \varepsilon \Rtilone, \quad \Rhattwo = 1 - \varepsilon \Rtiltwo, \quad \Rhatb = 1 -\varepsilon\Rtilb, \quad \that = \varepsilon \ttil. 
\end{equation}
A constant small-time velocity is assumed for the two interfaces, such that $\d \Rtilone/ \d \ttil = P$ and $\d \Rtiltwo/ \d \ttil = Q$. As a result, the initial positions and velocities of the two interfaces can be expressed as follows: 
\begin{subequations} \label{eq:3p-smalltime-pos-vel}
\begin{align}
   &\Rhatone(\that)  = 1 - \varepsilon P\ttil = 1 - P\that,  \\
   &\Rhattwo(\that)  = 1 - \varepsilon Q\ttil = 1 - Q\that,  \\
   &\uhatone  = \dd{\Rhatone}{\that}  = -P, \\
   &\uhattwo  = \dd{\Rhattwo}{\that}  = -Q.
\end{align}
\end{subequations}
Small-time radii $\Rtilone$, $\Rtiltwo$, and $\Rtilb$ are related to each other via Eq.~\eqref{eq:3P-Rb-ndim}. Following steps similar to those used to arrive at Eq.~\eqref{eq:2P-Rbtil}, $\Rtilb$ for three-phase reads as
\begin{equation}
\Rtilb = a \Rtilone + (1-b)\Rtiltwo. \label{eq:3P-Rbtil}
\end{equation}

\noindent According to the magnitude of the liquid-vapor and solid-vapor density ratio $\rrholv$ and $\rrhosv$, respectively, the vapor domain velocity $\uhatv$ (Eq.~\eqref{eq:uhatv(r,t)}) scales as 
\begin{subequations} \label{eq:3P-uv-scaling}
  \begin{align}
\uhatv &\approx  \mathcal{O}(1), & \text{[low density ratio scaling]} \label{eq:3P-uv-ldr-scaling} \\
\uhatv &\approx  \mathcal{O}\left(\frac{1}{\varepsilon}\right), & \text{[high density ratio scaling]} \label{eq:3P-uv-hdr-scaling}
\end{align}
\end{subequations}
in which the large liquid-vapor (similarly solid-vapor) density ratio is scaled as $\rrholv = \rrholvtil/\varepsilon$ with $\rrholvtil \sim \mathcal{O}(1)$. Therefore, we obtain two different small-time solutions for low and high liquid-vapor/solid-vapor density ratios.

\subsubsection{Small-time low-density-ratio solution (LDRS) to the three-phase spherical Stefan problem} \label{sec_3p-ldrs}

Proceeding analogously and assuming low density ratio between the three phases, the leading-order terms for the temperature equation in the three phases read as
\begin{subequations} \label{eq:3p-ld-smalltime-PDEs}
\begin{align}
\frac{\partial^2 \Ttils}{\partial \rtil^2} &= 0, && \in \Omegas \\
\frac{\partial^2 \Ttill}{\partial \rtil^2} &= 0, && \in \Omegal \\
\frac{\partial^2 \Ttilv}{\partial \rtil^2} &= 0. && \in \Omegav
\end{align}
\end{subequations}
The small-time solution for temperature in the solid domain corresponds to Eq.~\eqref{eq:2P-smalltime-Ts}. For the small-time liquid and vapor temperature solutions, $\Ttill(\rtil,\ttil)= D_1(\ttil)\,\rtil + D_2(\ttil)$ and $\Ttilv(\rtil,\ttil)= E_1(\ttil)\,\rtil + E_2(\ttil)$, respectively, the unknown functions (of time) $D_1$ and $D_2$, and $E_1$ and $E_2$ are determined by imposing appropriate boundary conditions. These are $\Ttill(\Rtilone,\ttil) = \Ttilim$, \; $\Ttill(\Rtiltwo,\ttil) = \Ttiliv$, \; $\Ttilv(\Rtiltwo,\ttil) = \Ttiliv$, and $\Ttilv(\Rtilb,\ttil) = \Thatinf$. Here, the leading-order interface temperatures are denoted $\Ttilim = \Thatm - \Gammam$ and  $\Ttiliv = \That_v - \Gammav$. Solving for the unknowns, the small-time temperature solution in the three phases (in terms of non-dimensional variables) reads as
\begin{subequations} \label{eq:3p-smalltime-sols}
\begin{align}
\Thats(\rhat, \that) &= \Ttilim, && 0\le \rhat<\Rhatone(\that) \\
\Thatl(\rhat, \that) &= \frac{\Ttiliv - \Ttilim}{\Rhattwo - \Rhatone}(\rhat - \Rhattwo) + \Ttiliv, && \Rhatone(\that)<\rhat<\Rhattwo(\that) \\
\Thatv(\rhat, \that) &= \frac{\Thatinf - \Ttiliv}{(a-b) + b\Rhattwo - a\Rhatone}(\Rhattwo - \rhat) + \Ttiliv. && \Rhattwo(\that)<\rhat \le \Rhatb(\that)  
\end{align}
\end{subequations}

The small-time solutions written is Eqs.~\eqref{eq:3p-smalltime-sols} are complete once the values of $P$ and $Q$ are determined. This can be achieved by substituting the small-time solutions into the Stefan Eqs.~\eqref{eq:Stefan-2P-ndim} and~\eqref{eq:Stefan-3P-ndim} for the melt and boiling fronts, respectively. Thus, we have the following two coupled equations in $P$ and $Q$, which are solved numerically:
\begin{subequations} \label{eq:3P-PQsystem}
\begin{align}
\frac{\Ttiliv-\Ttilim}{(P-Q)\that} &=\rrhosl\,\betam\left[1+\gammam\Ttilim-\frac{1}{2}\bigl(1-(\rrhosl)^2\bigr)\deltam\,P^2\right]P, \label{eq:3P-Psol}\\
\frac{\Ttiliv-\Ttilim}{(P-Q)\that}+\,\frac{\kappa^{\rm VL}(\Thatinf-\Ttiliv)}{\big(aP-b Q\big)\that} &= \betav\Big(-Q+(1-\rrhosl)P\Big)\,\bv.\label{eq:3P-Qsol}
\end{align}
\end{subequations}
$\bv$ in Eq.~\eqref{eq:3P-Qsol} is evaluated using small-time interface speeds $\uhatone = -P$ and $\uhattwo = -Q$ and interface temperature $\Thativ=\Ttiliv$ as
\begin{alignat}{2}\label{eq:bv0-def}
    \bv &= \Bigg[1+\gammav \Ttiliv-\frac{1}{2}(1-\rrhosl)^2\deltav\,P^2+\frac{1}{2}\deltav\Big\{(\rrholv-\rrhosv)P+(1-\rrholv)Q\Big\}^2 - \nonumber\\ 
    &\quad\quad \deltav\Big\{(\rrholv-\rrhosv)P+(1-\rrholv)Q-(1-\rrhosl)P\Big\}Q\Bigg].
\end{alignat}

Using the numerical solution of Eqs.~\eqref{eq:3P-PQsystem}, the small-time prediction of melt and boiling front positions are $\Rhatone = 1 - P\that_{\rm init}$ and $\Rhattwo = 1 - Q\that_{\rm init}$, respectively, and that of the outer free boundary is $\Rhatb = 1 - (aP + (1-b)Q)\that_{\rm init}$. A more accurate prediction of the initial outer boundary location can be obtained from the small-time estimates of $\Rhatone$ and $\Rhattwo$ through the use of Eq.~\eqref{eq:3P-Rb-ndim}: $\Rhatb = \sqrt[3]{a(\Rhatone^3 - 1) - b(\Rhattwo^3 - 1) + \Rhattwo^3}$. For a physically correct small-time solution, these predictions should satisfy the condition $\Rhatone < \Rhattwo < \Rhatb$. Furthermore, if Eq.~\eqref{eq:3P-Rb-ndim} is used then $\Rhatb \in \mathbb{R}$. For the three-phase case of Sec.~\ref{sec_3pstefan}, our empirical tests show that as long as $ \rhol/10 \lesssim \rhov \leq \rhol$, we obtain physically correct small-time predictions of the interface positions. Additionally, the two estimates of $\Rhatb$ are quite close to each other for this range of $\rhov$. Assuming a low density ratio, Myers et al.~\cite{myers2020stefan_varprops} introduced the small-time solution technique in the context of two-phase melting of a nanosphere induced by a convective boundary condition at the exterior boundary.  The authors considered tin and gold nanoparticles with a small solid-to-liquid density ratio of $\rrhosl = 1.0287$ and $\rrhosl = 1.1156$, respectively. To obtain physically correct small-time solutions for large density ratios ($\rrhosl > 10$ or $\rrholv > 10$) we use the correct scaling for the liquid and vapor phase velocity as written in Eqs.~\eqref{eq:2P-ul-hdr-scaling} and~\eqref{eq:3P-uv-hdr-scaling}, respectively. This is analyzed in the next section.

\subsubsection{Small-time high-density-ratio solution (HDRS) to the three-phase spherical Stefan problem} \label{sec_3p-hdrs}

If the density ratio between the three phases is large, then the leading order terms for the temperature equation in the three phases read as
\begin{subequations} \label{eq:3p-hd-smalltime-PDEs}
\begin{align}
\frac{\partial^2 \Ttils}{\partial \rtil^2} &= 0, && \in \Omegas \\
\frac{\partial^2 \Ttill}{\partial \rtil^2} &= M\frac{\partial \Ttill}{\partial \rtil}, && \in \Omegal \\
\frac{\partial^2 \Ttilv}{\partial \rtil^2} &= N\frac{\partial \Ttilv}{\partial \rtil}, && \in \Omegav
\end{align}
\end{subequations}
in which $M$ and $N$ are $\mathcal{O}(1)$ constants. Proceeding analogously to Sec.~\ref{sec_2p-hdrs} and imposing appropriate boundary conditions on the three interfaces, the small-time HDRS for temperature in the three phases reads as

\begin{subequations}\label{eq:3o-hd-smalltime-sols}
    \begin{align}
\Thats(\rhat, \that) &= \Ttilim, && 0\le \rhat<\Rhatone(\that) \\
\Thatl(\rhat, \that) &= \Ttiliv - \big(\Ttiliv - \Ttilim\big) \frac{(1-\Rhattwo)^2-(1-\rhat)^2}{(1-\Rhattwo)^2 - (1-\Rhatone)^2} , && \Rhatone(\that)<\rhat<\Rhattwo(\that) \\
\Thatv(\rhat, \that) &= \Thatinf- \big(\Thatinf - \Ttiliv\big) \frac{\big\{a(1-\Rhatone) + (1-b)(1-\Rhattwo)\big\}^2-(1-\rhat)^2}{\big\{a(1-\Rhatone) + (1-b)(1-\Rhattwo)\big\}^2 - (1-\Rhattwo)^2} . && \Rhattwo(\that)<\rhat \le \Rhatb(\that)  
\end{align}
\end{subequations}
The small-time HDRS for temperature written in Eqs.~\eqref{eq:3o-hd-smalltime-sols} is substituted into the Stefan Eqs.~\eqref{eq:Stefan-2P-ndim} and~\eqref{eq:Stefan-3P-ndim} to obtain the system of equations in $P$ and $Q$:
\begin{subequations}\label{eq:3p-hd-PQsystem}
    \begin{align}
\frac{2P(\Ttiliv-\Ttilim)}{(P^2-Q^2)\that} &=\rrhosl\,\betam\left[1+\gammam\Ttilim-\frac{1}{2}\bigl(1-(\rrhosl)^2\bigr)\deltam\,P^2\right]P, \label{eq:3P-hd-Psol}\\
\frac{2Q(\Ttiliv-\Ttilim)}{(P^2-Q^2)\that} +\, \frac{2Q\kappa^{\rm VL}(\Thatinf-\Ttiliv)}{\big(\big(aP +(1-b)Q\big)^2-Q^2\big)\that} &= \betav\Big(-Q+(1-\rrhosl)P\Big)\,\bv.\label{eq:3P-hd-Qsol}
\end{align}
\end{subequations}
Here, $\bv$ is given by Eq.~\eqref{eq:bv0-def}

Through numerical experiments we find that the solution to Eqs.~\eqref{eq:3p-hd-PQsystem} yields physically correct small-time positions of the interfaces that satisfy the condition $\Rhatone < \Rhattwo < \Rhatb$. This is even for vapor density as low as $\rhov \approx \rhol/1000$.

\section{A fixed-grid sharp-interface technique to solve the three-phase Stefan problem numerically} \label{sec_sharp_interface_method}

\begin{figure}[]
	\begin{center}
   \includegraphics[scale=0.44]{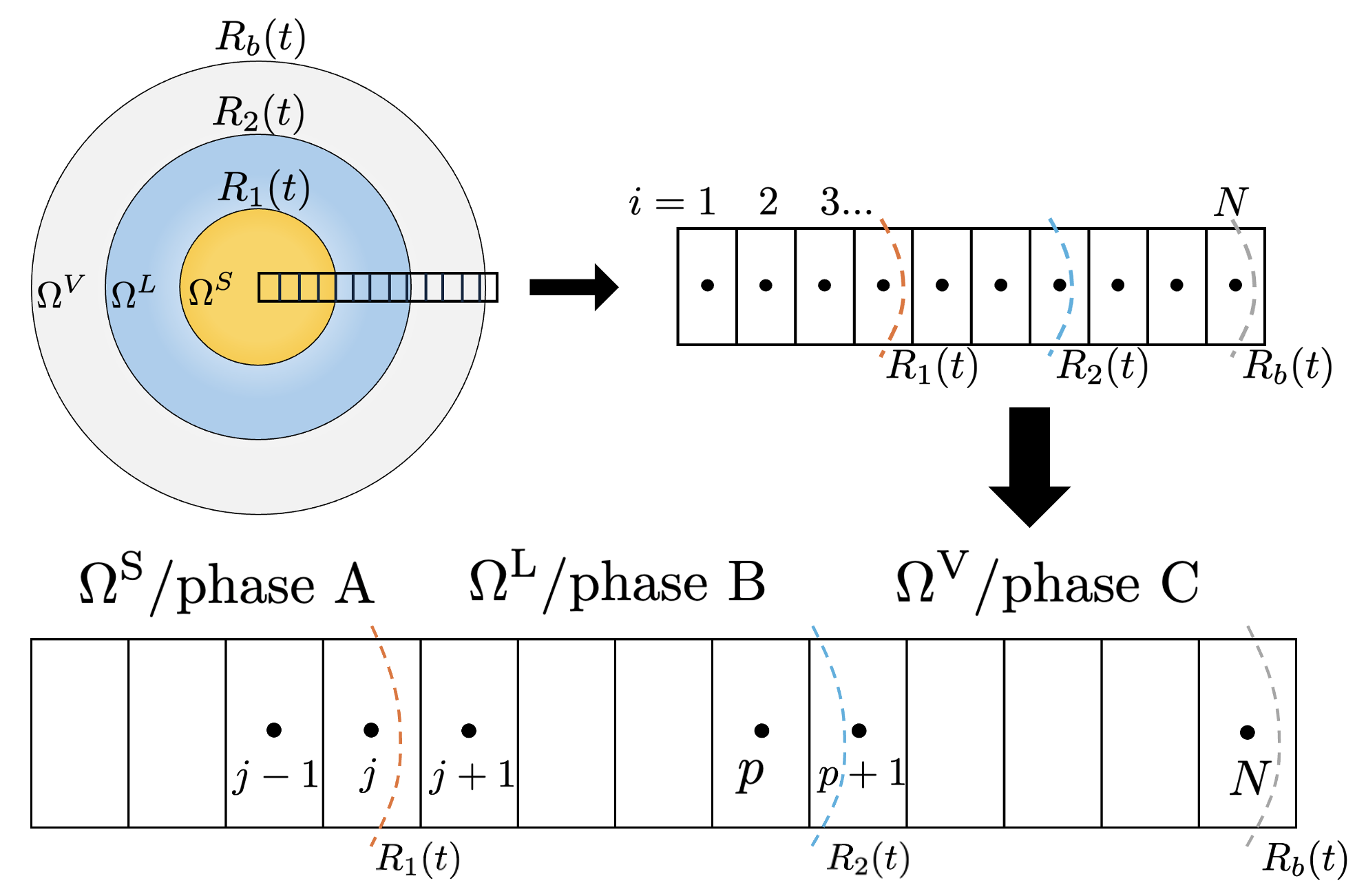} \label{fig_SharpIntffig1}
	\end{center}
	\caption{Schematic of the 1D grid used to discretize the spherical heat equations using the sharp interface technique. Irregular cells $j$ and $j+1$, $p$ and $p+1$, and $N$ abut  interfaces separating phases $A$ and $B$, and phases $B$ and $C$, and outer free boundary, respectively. The cell centers are marked with $(\bullet)$, and the evolving interfaces and free boundary are marked with dashed lines (\texttt{---}). }\label{fig_grid_sharp_interface}
\end{figure}

In this section, we present a fixed-grid, sharp-interface method to solve the dimensionless heat/temperature Eqs.~\eqref{eq:3P-nondim-S}-\eqref{eq:3P-nondim-V} in the evolving solid, liquid and vapor domains. Stefan conditions and interface temperature conditions on the evolving solid-liquid and liquid-vapor interfaces are imposed sharply in this approach. We describe the discretization process for the three-phase case as it is more general. In the following discussion we drop the $\widehat{(.)}$ symbol from the dimensionless variables for the ease of notation.  

The one-dimensional domain $\Omega: = 0 \le r \le \Rmax$ is discretized into $\Nmax$ cells of uniform size $\Delta r$, such that $\Nmax = \lceil \Rmax/\Delta r \rceil$. Here, $\Rmax = (\rrhosv)^{\frac{1}{3}}$ is the maximum radius of the sphere, when all of the initial solid particle (of unit initial radius) has evaporated.  The discrete temperature $T_i$ and velocity $u_i$ are both stored at the $i^{\rm th}$ (with $i = 1,\ldots, N$) cell-center having a coordinate $r_i = (i - \half) \Delta r$. $\Nmax$ number of cells are computed based on user-defined $\Nmin$ number of cells that are needed to resolve the initial thickness of the liquid and vapor layers, $\Delta_0^{\rm L}$ and $\Delta_0^{\rm V}$, respectively. Specifically, we define $\Delta r := \textrm{min}(\Delta_0^{\rm L},\Delta_0^{\rm V})/\Nmin$ to control the grid resolution. We compute the initial liquid and vapor layers' thicknesses by $\Delta_0^{\rm L} = \Rtwo - \Rone = (P - Q)t_{\rm init}$ and $\Delta_0^{\rm V} = \Rb - \Rtwo =   ( bQ - aP)t_{\rm init}$. 

The heat equation is discretized using a second-order finite difference stencil. We treat conduction/diffusion implicitly and convection explicitly. The heat equation is discretized only for regular grid cells that do not contain or abut the moving interfaces ($s_1 = \Rone$, $s_2 = \Rtwo$, and $s_3 = \Rb$). In contrast, irregular cells containing or abutting the interface impose interface temperature conditions. With $n$ denoting the time level, and $\Delta t$ denoting the time-step size, the discrete form of the heat equation for a regular grid cell $i$ at (midpoint) time $t = (n+\half)\Delta t$ reads as
\begin{align} \label{eqn_discrete_heat_reg}
\frac{T_i^{n+1} - T_i^n}{\Delta t} + \bigg(u_i \frac{\widebar{T}_{i+\half} - \widebar{T}_{i-\half}}{\Delta r} \bigg)^{n+\half}  = \frac{\alpha_i}{r_i^2} \frac{1}{\Delta r}\bigg[ & r_{i+\half}^2 \frac{T^{n+1}_{i+1} - T^{n+1}_{i}}{2\Delta r} - r_{i-\half}^2 \frac{T^{n+1}_{i} - T^{n+1}_{i-1}}{2\Delta r} \nonumber  \\
& + r_{i+\half}^2 \frac{T^{n}_{i+1} - T^{n}_{i}}{2\Delta r} - r_{i-\half}^2 \frac{T^{n}_{i} - T^{n}_{i-1}}{2\Delta r}\bigg].
\end{align}
Here, $\widebar{T}$ denotes the CUI (cubic upwind interpolated)~\cite{nangia2019robust} limited temperature field on the left ($i - \half$) and right ($i+\half$) faces of the regular grid cell $i$. Because we store the discrete values of temperature at cell centers, the discretized form of the Laplace operator in spherical coordinates does not suffer from the $r = 0$ singularity. For the first grid cell with index $i = 1$, we directly impose the homogeneous Neumann boundary condition on its left face.  An extrapolation of the form $\phi^{n+\half} = \threehalf \phi^n - \half \phi^{n-1}$ is performed to evaluate the velocity and temperature at time level $n+\half$ for the convective term. As the simulation advances, the number of unknown discrete temperatures $N$ varies based on the position of the outer boundary $s_3 = \Rb(t)$. We do not solve for temperature locations beyond $\Rb(t)$ as those are assumed to be at a constant value of $\Tinf$. We stop the simulation when the solid is almost completely melted, but has not yet fully evaporated. Therefore, at any instant the number of unknowns (equivalently the number of cells) $N$ in the system of equations are less than $\Nmax$.  

Grid cells adjacent to the interface, say $j$ and $j+1$, are indicated as irregular in Fig.~\ref{fig_grid_sharp_interface}.  We do not discretize the heat equation for irregular cells. To complete the system of equations for $N$ unknown temperatures, we need additional equations for the irregular cells. One-sided quadratic extrapolation is used to calculate the temperature at the interface by associating cells $j-2$, $j-1$, and $j$ with phase $A$ and cells $j+1$, $j+2$, and $j+3$ with phase $B$. The one-sided extrapolated temperature at the interface is equated to the interface temperature $T^{\rm I}$ (i.e., $T_m^{\rm I}$ or $T_v^{\rm I}$) to close the system of equations:
\begin{subequations} \label{eqn_extrap_temp}
 \begin{alignat}{2}
 &T_A \bigg|_{r = s} =  a_1 T_{A,j} + a_2 T_{A,j-1} + a_3 T_{A,j-2} = T^{\rm I}, \\
 &T_B \bigg|_{r = s} =  b_1 T_{B,j+1} + b_2 T_{B,j+2} + b_3 T_{B,j+3} = T^{\rm I}.
 \end{alignat}
\end{subequations}
Similarly, for the outer boundary at $s_3 = \Rb$, we impose the temperature boundary condition using the three neighboring temperatures in phase C as
\begin{equation} \label{eqn_extrap_temp_Rb}
 T_C \bigg|_{r = s_3} =  c_1 T_{C,N} + c_2 T_{C,N-1} + c_3 T_{C,N-2} = T_\infty.
 \end{equation}
 
The coefficients $a_k$, $b_k$, and $c_k$ appearing in Eqs.~\eqref{eqn_extrap_temp} and~\eqref{eqn_extrap_temp_Rb} are determined by fitting a one-dimensional, second-order Lagrange polynomial to the neighboring nodes. Phase $A$, for example, has coefficients $a_k$ as follows:
\begin{subequations}\label{eqn_coef_ak}
\begin{alignat}{2}
	&	a_1=\frac{\left(s - r_{j-1}\right)\left(s - r_{j-2}\right)}{\left(r_j-r_{j-1} \right)\left(r_{j} -r_{j-2}\right)}, \\
		&  a_2=\frac{\left(s-r_j\right)\left(s-r_{j-2}\right)}{\left(r_{j-1}-r_j\right)\left(r_{j-1}-r_{j-2}\right)}, \\
		&  a_3=\frac{\left(s-r_j\right)\left(s-r_{j-1}\right)}{\left(r_{j-2}-r_j\right)\left(r_{j-2}-r_{j-1}\right)}.
\end{alignat}
\end{subequations}
A similar process is followed for irregular cells $p$ and $p+1$ that abut the second interface, and cell $N$ that abuts the outer boundary. 

The discrete Eqs.~\eqref{eqn_discrete_heat_reg}, \eqref{eqn_extrap_temp}, and~\eqref{eqn_extrap_temp_Rb} form a system of $N$ equations that are inverted using a sparse direct solver to calculate the $N$ unknown temperatures. At this stage, only the velocities of the two interfaces, $u_1^{n+1}$ and $u_2^{n+1}$ need to be determined. Calculating them requires evaluating the one-sided derivatives of temperature $T^{n+1}$ required for the right hand side of Stefan Eqs.~\eqref{eq:Stefan-2P-ndim} and~\eqref{eq:Stefan-3P-ndim}. A one-sided derivative of temperature at an interface may be expressed in terms of neighboring nodal temperatures in phase $A$ as follows:
\begin{subequations} \label{eqn_taylor_series_dtdx}
\begin{alignat}{2} 
    \dd{T_A}{r}\Bigg|_{r = s} &=  \sum_{j = 1}^{3}c_{A,j} T_{A,j}, \label{eqn_taylor_series_dtdxa}\\
    \hookrightarrow \dd{T_A}{r}\Bigg|_{r = s} &= \sum_{j = 1}^{3} c_{A,j}\left(T_A\bigg|_s + \dd{T_A}{r}\bigg|_s (r_j - s) + \half \dD{T_A}{r}\bigg|_s (r_j - s)^2 + \mathcal{O}\left((r_j - s)^3\right) \right). \label{eqn_taylor_series_dtdxb} 
\end{alignat}
\end{subequations}
Eq.~\eqref{eqn_taylor_series_dtdxb} follows from Eq.~\eqref{eqn_taylor_series_dtdxa} through the use of the Taylor series expansion to express neighboring nodal temperatures in terms of the interface temperature and its derivatives. Ignoring higher order derivatives in Eq.~\eqref{eqn_taylor_series_dtdx}, and equating coefficients of $T_A$, $\dd{T_A}{r}$, and $\dD{T_A}{r}$ on both sides of the equation, yields a $3\times3$ system of equations for the coefficients $c_{A,j}$
 \begin{equation}
 	\begin{bmatrix}
 		1 & 1 & 1 \\
 		\Delta r_1 & \Delta r_2 & \Delta r_3 \\
 		(\Delta r_1)^2 & (\Delta r_2)^2 & (\Delta r_3)^2 \\
 	\end{bmatrix}
 	\begin{bmatrix}
 		c_{A,1} \\
 		c_{A,2} \\
 		c_{A,3} \\
 	\end{bmatrix}
 	=
 	\begin{bmatrix}
 		0 \\
 		1  \\
 		0 \\
 	\end{bmatrix}.
 \end{equation}
 Here, $\Delta r_j = (r_j - s)$. To calculate $\dd{T_B}{r}\bigg|_{r = s}$ for phase $B$, a similar procedure is followed. Once the positions of melt and boiling fronts are updated using $u_1^{n+1}$ and $u_2^{n+1}$, the position of the outer boundary is updated using Eq.~\eqref{eq:3P-Rb-ndim}. 

\section{Software implementation}
All MATLAB- and Mathematica-based codes used to generate results in this work can be found on the GitHub repository~\url{https://github.com/amneetb/SphericalStefan}.

\section{Results}\label{sec_results}

We present results for both the two- and three-phase spherical Stefan problems. While the main focus is on the latter problem, the former problem is solved for validating the numerical method against the results of Font et al.~\cite{font2015nanoparticle}. 

\subsection{Two-phase spherical Stefan problem}\label{sec_2P_Stefan_results}

\begin{table}[]
\centering
\caption{Thermophysical properties of gold used to simulate the two-phase spherical Stefan problem.}
\label{tab_thermophys_properties_Au}
\begin{tabular}{ll l}
Property & Value & Units\\
\midrule
Density of solid $\rhos$                  & 19300  & kg/m$^3$ \\
Density of liquid $\rhol$                 & 17300  & kg/m$^3$ \\
Thermal conductivity of solid   $\ks$     & 317   & W/m.K  \\
Thermal conductivity of liquid  $\kl$     & 106   & W/m.K    \\
Specific heat of solid  $\cps$            & 129   & J/kg.K      \\
Specific heat of liquid  $\cpl$           & 163   & J/kg.K   \\
Melting temperature  $T_m$                & 1337  & K     \\
Reference temperature $T_r$               & 1337  & K \\   
Latent heat of melting $L_m$              & 63700  & J/kg    \\
Surface energy coefficient $\sigmasl$                 & 0.27  & N/m    \\
\bottomrule
 \end{tabular}
\end{table}


We simulate the melting of a gold nanoparticle of radius 100 nm using the same setup as Font et al.~\cite{font2015nanoparticle}.  The thermophysical properties of gold are written in Table~\ref{tab_thermophys_properties_Au} and are taken from references~\cite{font2015nanoparticle,buffat1976size}. The initially solid gold nanoparticle occupies the domain $\Omega : 0\le \rhat \le 1$. As the gold's solid-liquid density ratio is low ($\rrhosl = 1.1156$), the LDRS of Sec.~\ref{sec_2p-ldrs} is used to initialize temperature in the solid and liquid phases, along with interface positions and velocities. The small-time solution is evaluated at $\that_{\rm init} = 0.001$, and the simulation proceeds from there. 

\begin{figure}[]\label{fig_2PMyersComp}
	\begin{center}
        \subfigure[$\betam = 100$]{%
        \includegraphics[scale=0.21]{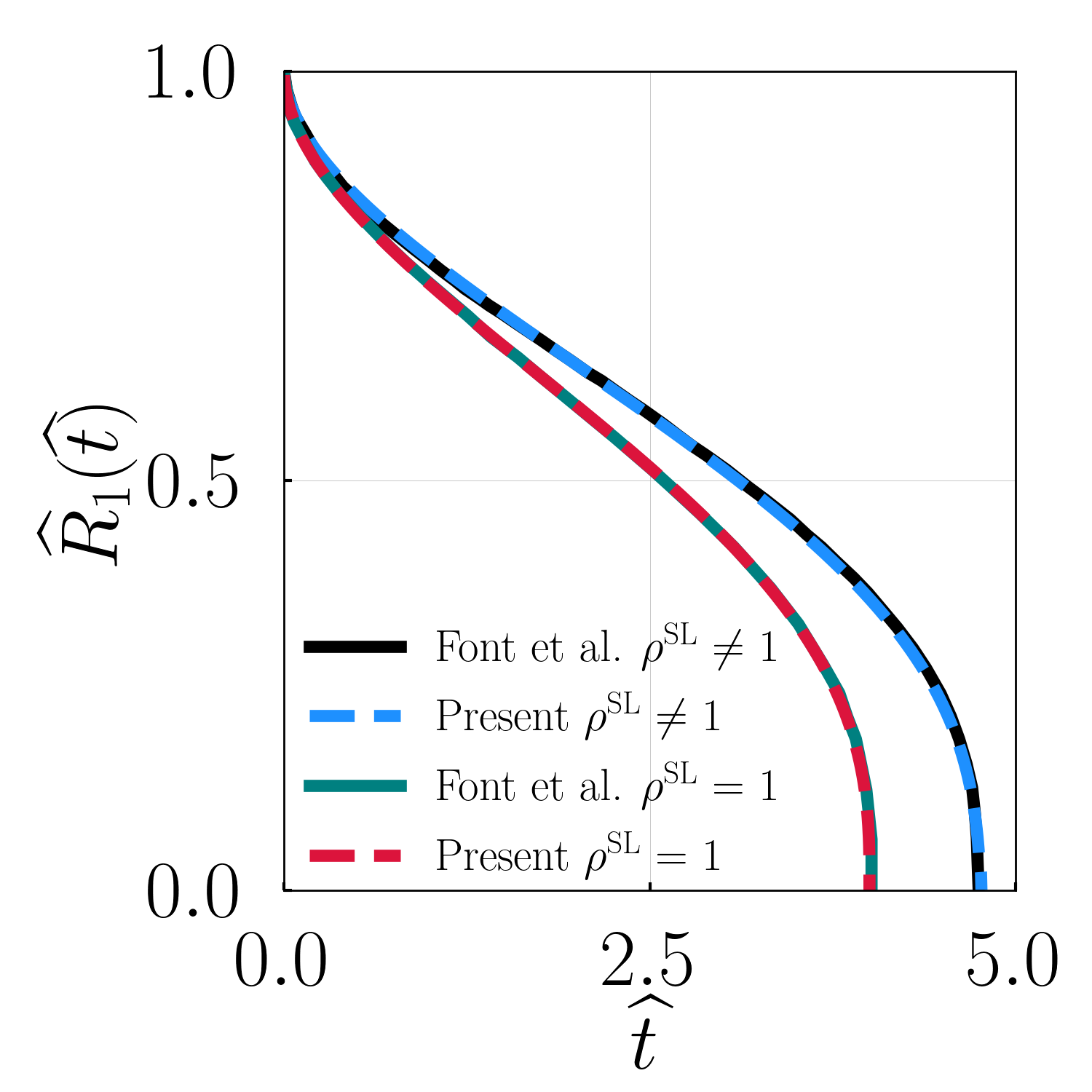}
        \label{fig_CompBeta100}
        }
        \subfigure[$\betam = 10$]{%
        \includegraphics[scale=0.21]{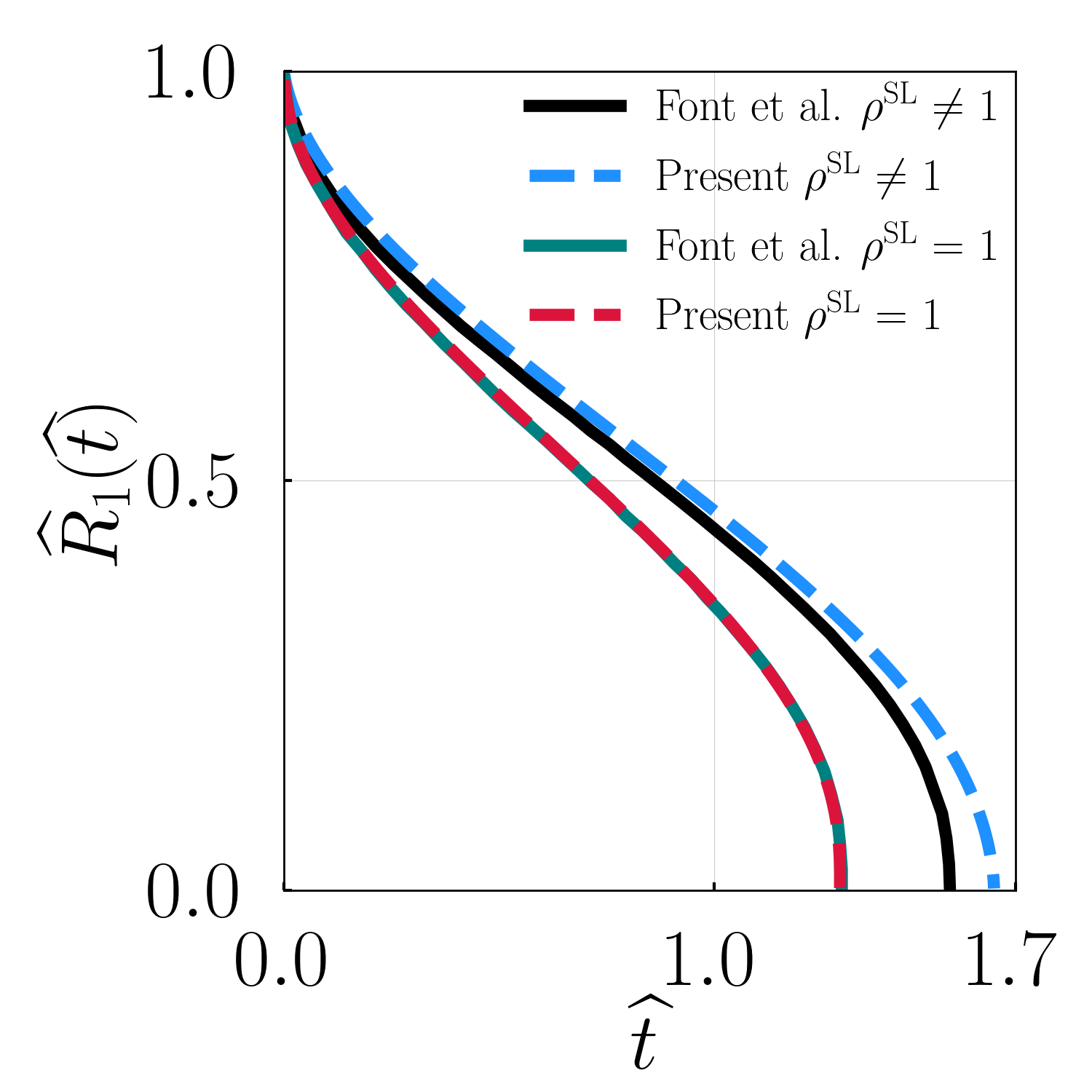}
        \label{fig_CompBeta10}
        }
        \subfigure[$\betam = 100$]{%
        \includegraphics[scale=0.25]{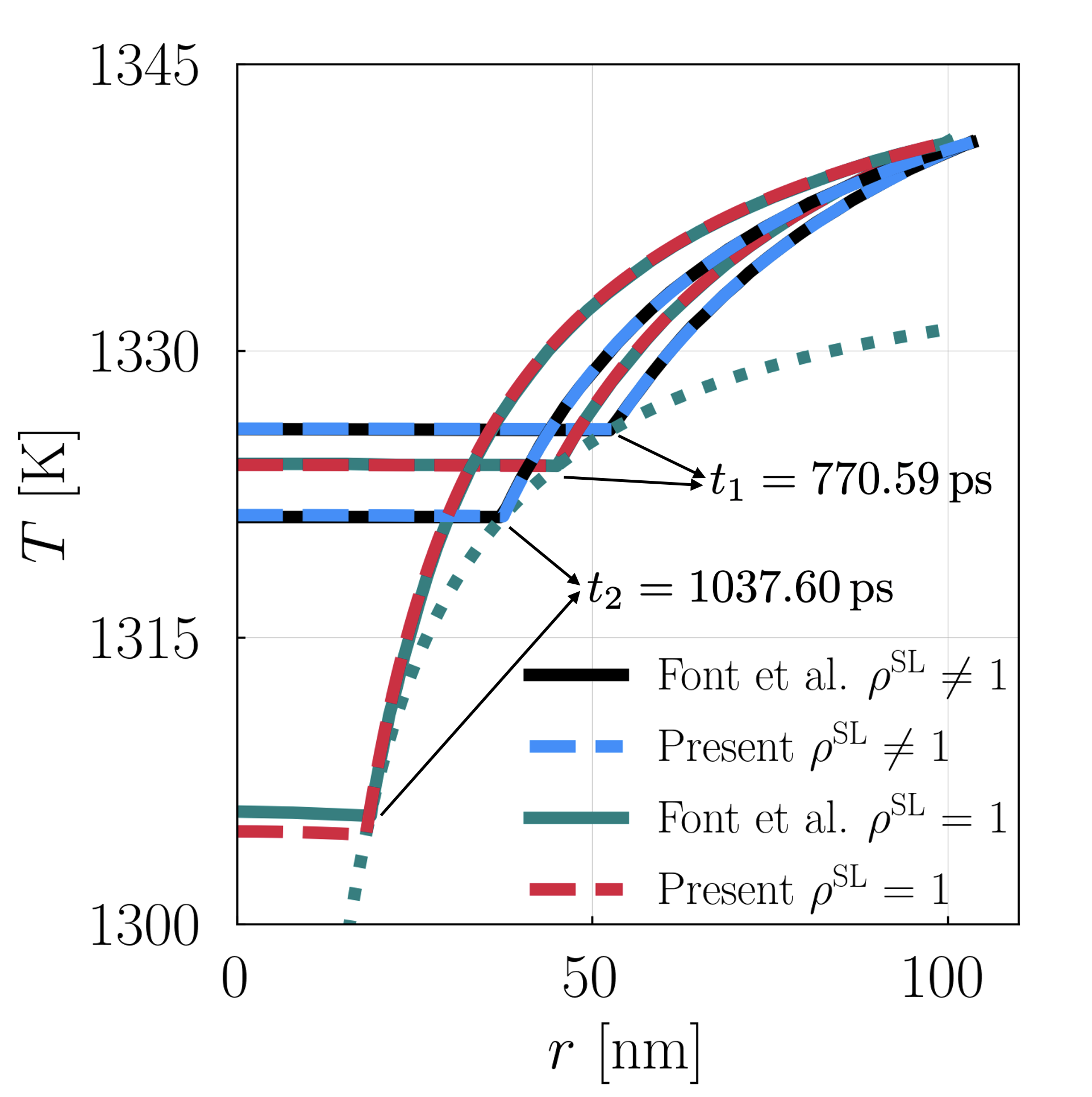}
        \label{fig_CompTemp}
        }
	\end{center}
\caption{Melting of a gold nanoparticle of initial radius $R_0 = 100$ nm for two Stefan numbers: $\betam = 100$ and $\betam = 10$. $\rrhosl = 1$ implies gold solid and liquid phases have the same density, and $\rrhosl \ne1$ implies different densities. \subref{fig_CompBeta100} Time evolution of the melt front for Stefan number $\betam = 100$ and \subref{fig_CompBeta10} for Stefan number $\betam = 10$. \subref{fig_CompTemp} Dimensional temperature variation within the nanoparticle at two different physical times for Stefan number $\betam = 100$. Results are compared against those reported in Font et al.~\cite{font2015nanoparticle}. The dotted line (\texttt{..}) shows the melt temperature variation.}
\end{figure}

Figs.~\ref{fig_CompBeta100} and~\ref{fig_CompBeta10} compare the time evolution of the melt front $\Rhatone$ obtained using our numerical method with that reported in Font et al.~\cite{font2015nanoparticle} for two Stefan numbers: $\betam = 100$ (slow phase change), and $\betam = 10$ (fast phase change). The external temperature at the free boundary can be defined by fixing the Stefan number $ \betam = L_m/(\cpl  \Delta T)$ and the reference temperature $T_r = T_m$. The time evolution of the melt front is also compared for matched ($\rhos = \rhol = 19300$ kg/m$^3$) and unmatched ($\rhos =19300$ kg/m$^3$, $\rhol = 17300$ kg/m$^3$) densities of gold solid and liquid phases. From Figs.~\ref{fig_CompBeta100} and~\ref{fig_CompBeta10}, it can be observed that our numerical results match quite well with those reported in~\cite{font2015nanoparticle}, with a small disagreement later for the $\beta_m = 10$ and $\rrhosl \ne 1$ case. To rule out any grid convergence issue for our simulations, we simulated the $\beta_m = 10$ and $\rrhosl \ne 1$ case using four grid sizes and obtained essentially the same converged solution; see~\ref{sec_2P_Stefan_Convergence}. It is also clear from Figs.~\ref{fig_CompBeta100} and~\ref{fig_CompBeta10} that the melt times differ between the matched and unmatched density cases. Considering $\betam = 100$, when the density change of gold is neglected during its melting, we get a melt time of $\that \approx 4$. When the density change is considered, the melt time is $\that \approx 4.7$, which is an approximate $15 \%$ increase. Particle melt times are longer due to the kinetic energy present in the system. The system loses a portion of its total energy (to the kinetic energy of the interface) for phase change as a result. The particle melt time is defined as the time when $\Rhatone$ abruptly approaches zero value. In Fig.~\ref{fig_CompTemp} we compare the dimensional temperature variation within the nanoparticle at two different physical times ($t$ = 770.59 ps and $t$ = 1037.6 ps) for the Stefan number $\betam = 100$. The dashed lines represent the reference solution taken from~\cite{font2015nanoparticle} and the dotted line shows the evolution of the interface temperature as its radius decreases over time. The decrease in interface temperature corresponds to the depression in melting temperature according to the Gibbs-Thompson relation. It can be observed that the temperature profiles agree quite well between the two studies.  

The results of this section validates our fixed-grid sharp-interface numerical method initialized using the small-time solution. 


\subsection{Three-phase spherical Stefan} \label{sec_3pstefan}

\begin{table}[]
\centering
\caption{Thermophysical properties of aluminum used to simulate the three-phase Stefan problem. The vapor density is artificially increased here due to the limitation of the small-time analysis. The remainder of the thermophysical properties of aluminum are physical and are taken from references~\cite{thirumalaisamy2023low,doble2007perry,hatch1984aluminium,desai1987thermodynamic}.}
\label{tab_thermophys_properties_Al}
\begin{tabular}{ll l}
Property & Value & Units\\
\midrule
Density of solid $\rhos$                  & 2698.72  & kg/m$^3$ \\
Density of liquid $\rhol$                 & 2368  & kg/m$^3$ \\
Density of vapor $\rhov$                  & \{500, 23\} & kg/m$^3$ \\
Thermal conductivity of solid   $\ks$     &   211  & W/m.K  \\
Thermal conductivity of liquid  $\kl$     & 91   & W/m.K    \\
Thermal conductivity of vapor  $\kv$      & 115.739  &W/m.K    \\
Specific heat of solid  $\cps$            & 910 & J/kg.K      \\
Specific heat of liquid  $\cpl$           & 1042.4 & J/kg.K   \\
Specific heat of vapor  $\cpv$            & 770.69 &  J/kg.K   \\
Melting temperature  $T_m$                & 933.6  & K     \\
Vaporization temperature  $T_v$           &  2767  & K     \\
Reference temperature  $T_r$              & 933.6  & K     \\
Latent heat of melting $L_m$              &  383840 & J/kg    \\
Latent heat of vaporization $L_v$         &  9462849.518 & J/kg    \\
Surface energy coefficient  $\sigmasl$    &  0.121 & N/m    \\
Surface energy coefficient  $\sigmalv$    &  0 & N/m    \\
\bottomrule
 \end{tabular}
\end{table}

The problem setup remains similar to the two-phase Stefan problem as described in Sec.~\ref{sec_2P_Stefan_results} with the exception of the material properties and externally imposed temperature $\Tinf$. For the three-phase spherical Stefan problem, aluminum's thermophysical properties are considered; see Table~\ref{tab_thermophys_properties_Al} for details. We consider two vapor densities: (i) $\rhov$ = 500 kg/m$^3$ and (ii) $\rhov$ = 23 kg/m$^3$ to test the small-time LDRS and HDRS of Sec.~\ref{sec_3P_small_time}, respectively. Though the HDRS discussed in Sec.~\ref{sec_3p-hdrs} can be used to initialize the simulation for even lower vapor densities, we limit the vapor density to $\rhov \approx \rhol/100$ to avoid extremely small time steps. At very low vapor densities, to conserve mass, the initial velocity in the vapor phase becomes extremely large. Accordingly, the time step size needs to be decreased significantly to satisfy the CFL condition and to maintain numerical stability. The outer boundary temperature is maintained at $\Tinf=4000~\mathrm{K}$. Because the temperature at the outer boundary is larger than aluminum's vaporization temperature ($T_v =2767~\mathrm{K}$),  the initially solid nanoparticle melts and boils simultaneously. The small-time solution is evaluated at $\that_{\rm init} = 0.001$, and the simulation proceeds from there. Temperatures in the solid, liquid, and vapor phases, along with interface positions and velocities, are set according to the small-time solution discussed in Sec.~\ref{sec_3P_small_time}.  

\begin{figure}[]
	\begin{center}
        \subfigure[]{\includegraphics[scale=0.18]{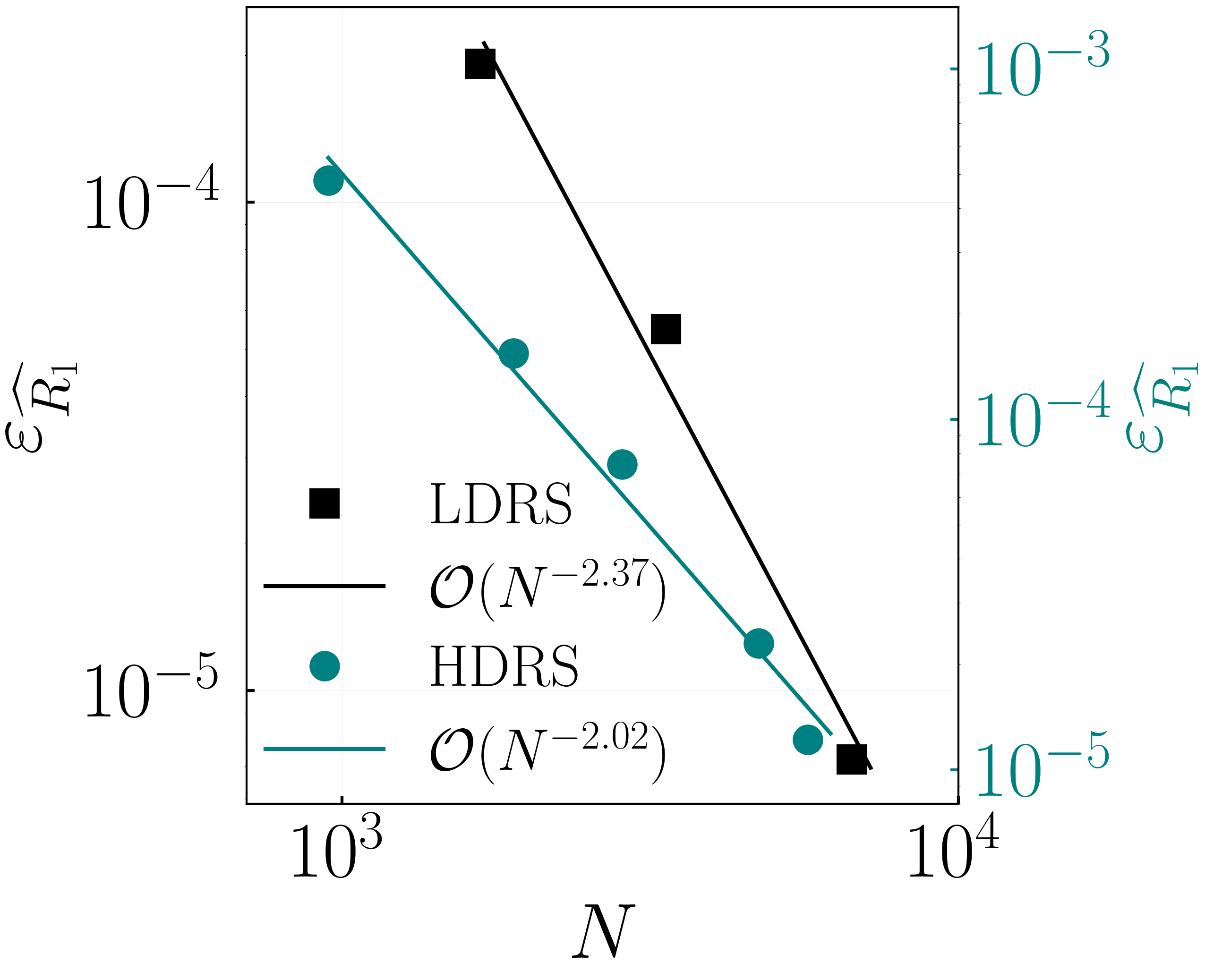}
        \label{fig_IntfConv3P}
        }
        \subfigure[]{\includegraphics[scale=0.18]{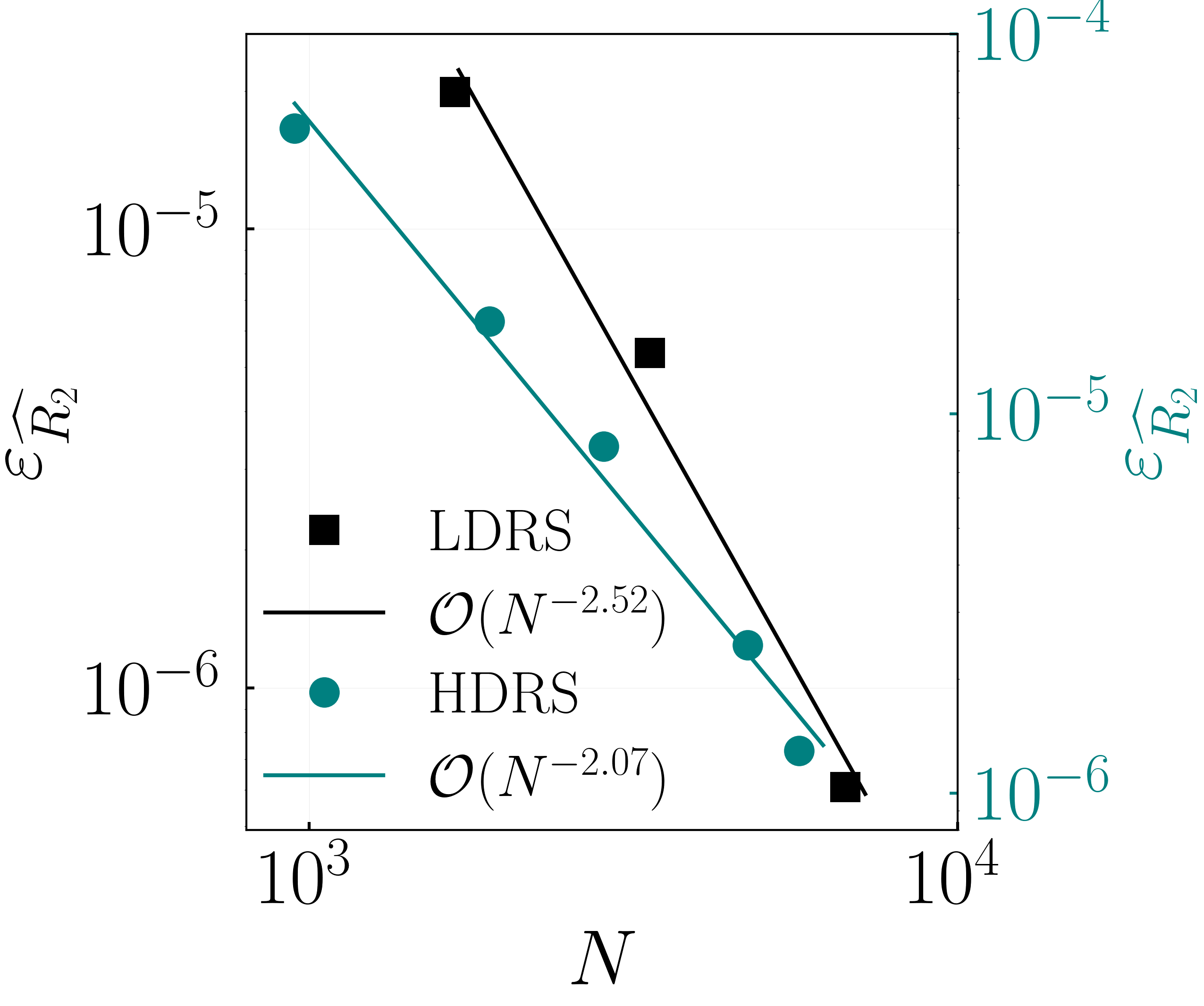}
        \label{fig_IntfR2Conv3P}
        }
        \subfigure[]{\includegraphics[scale=0.18]{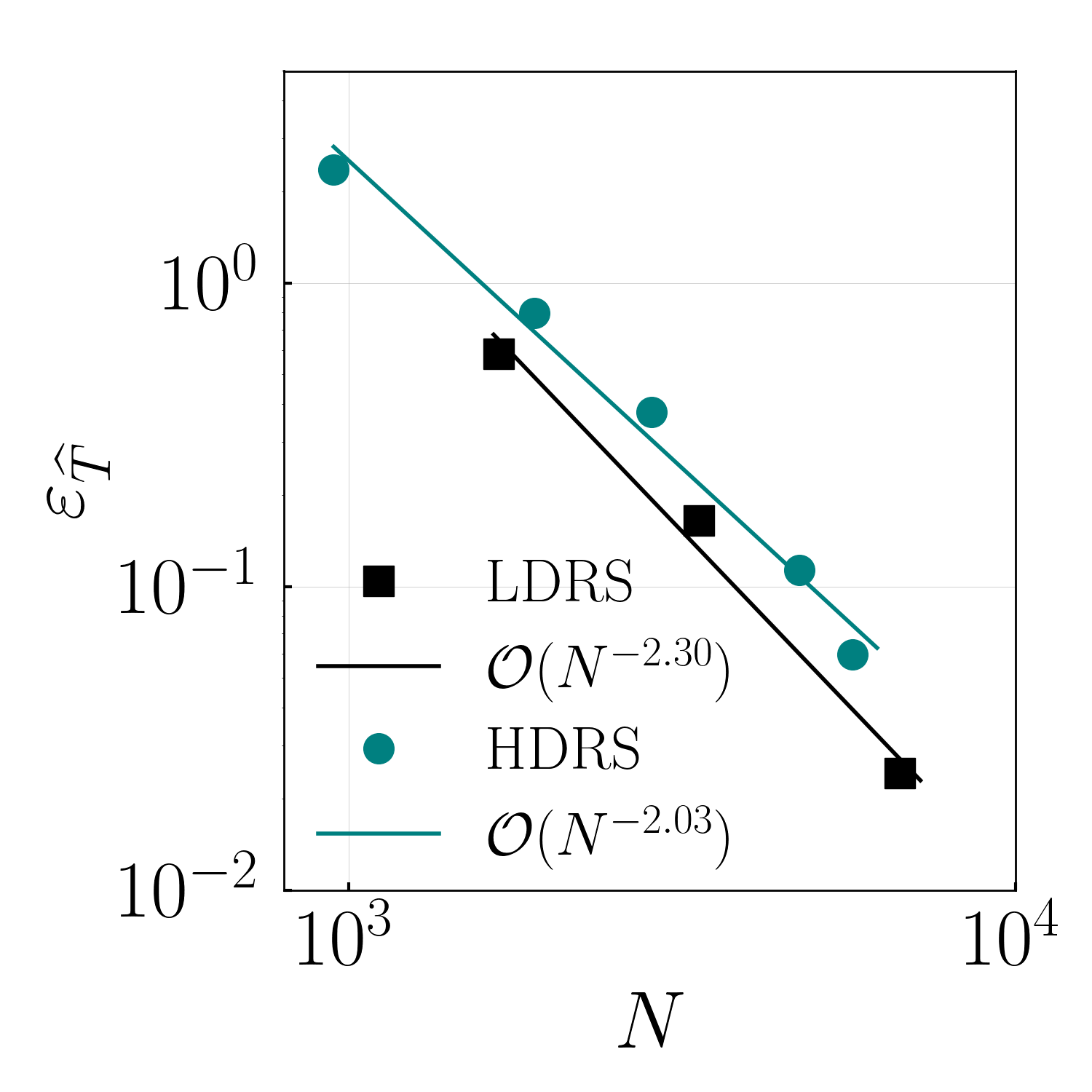}
        \label{fig_TempConv3P}
        }
	\end{center}
\caption{Convergence rate of the numerical error for \subref{fig_IntfConv3P} interface position $\Rhatone$, \subref{fig_IntfR2Conv3P} interface position $\Rhattwo$
and \subref{fig_TempConv3P}  temperature distribution as a function of grid size $N$. For interface positions, errors are computed over the time interval $\that = 0.1911$ to $\that = 0.1930$. For temperature, the error is computed at $\that = 0.19$.}
\end{figure}

\begin{figure}[]
	\begin{center}
        \subfigure[]{\includegraphics[scale=0.13]{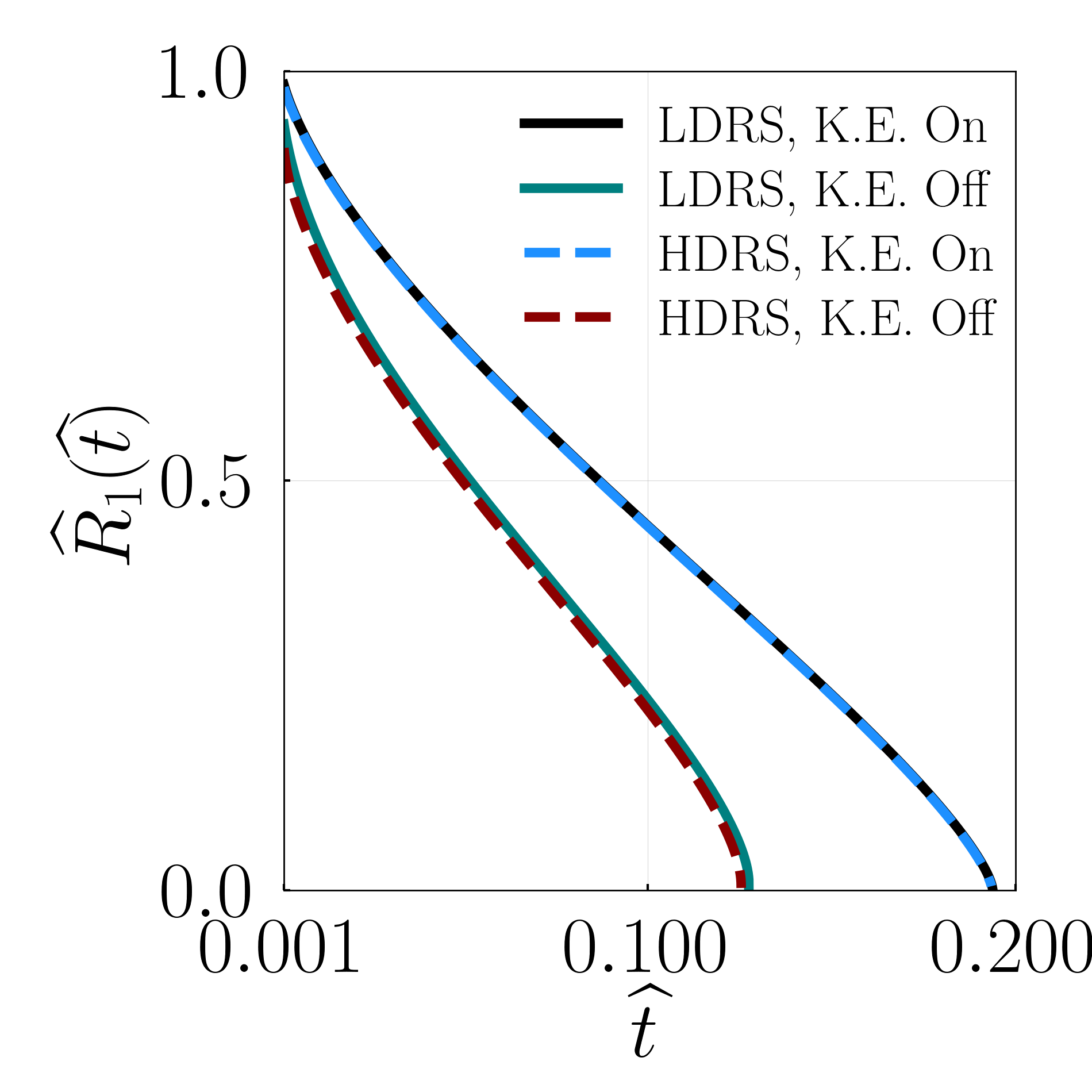}
        \label{fig_R1}
        }
        \subfigure[]{\includegraphics[scale=0.13]{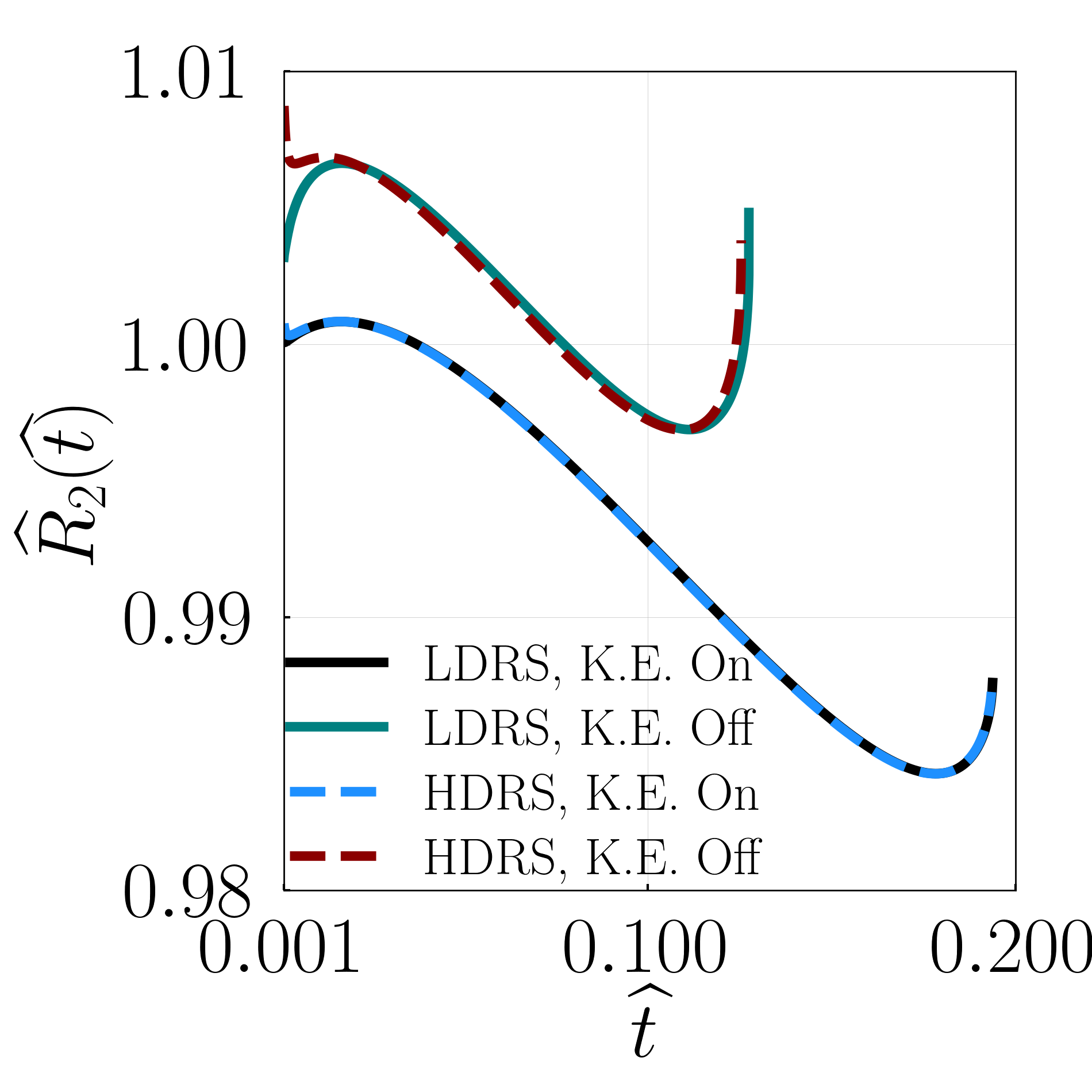}
        \label{fig_R2}
        }
        \subfigure[]{\includegraphics[scale=0.13]{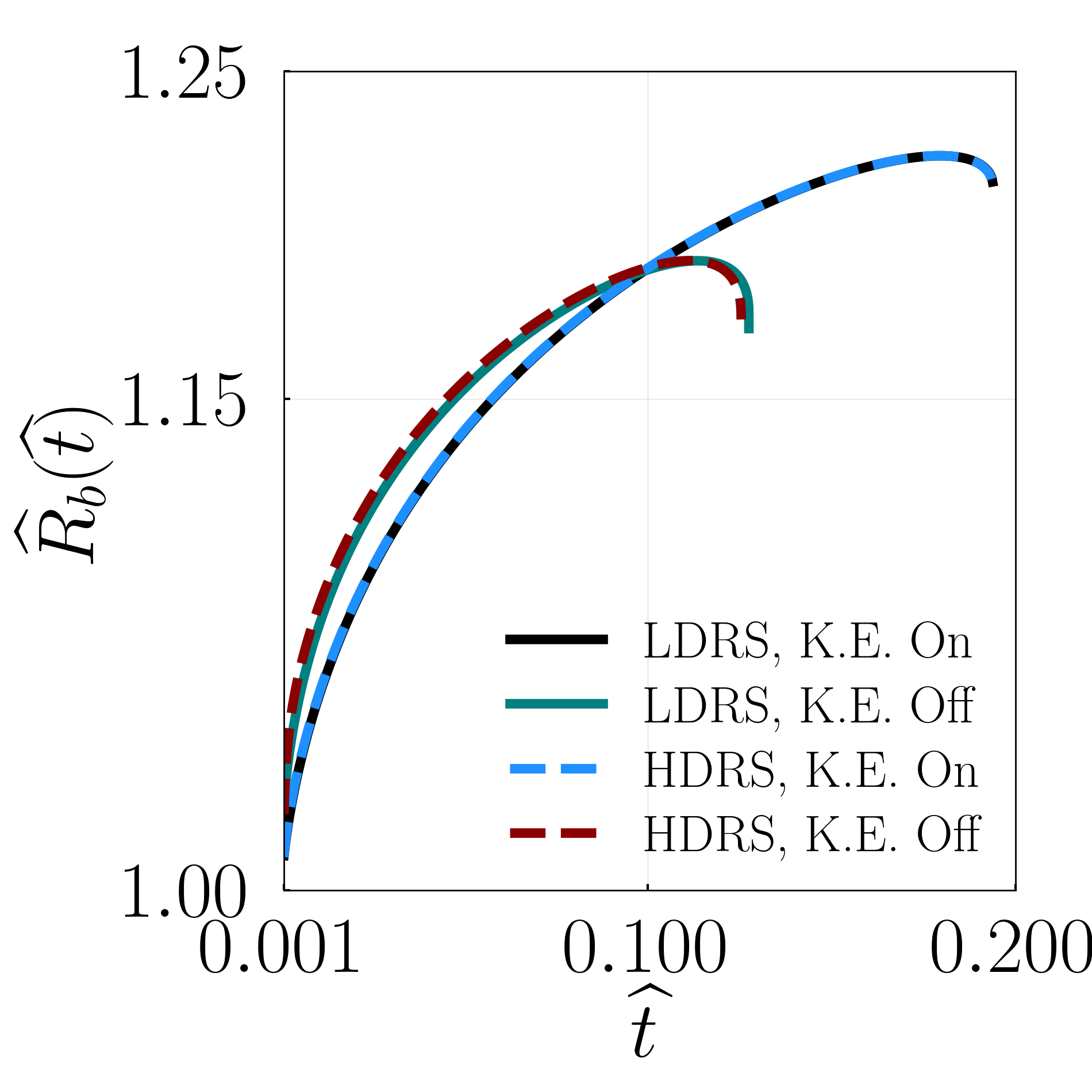}
        \label{fig_Rb}
        }
        \subfigure[]{\includegraphics[scale=0.24]{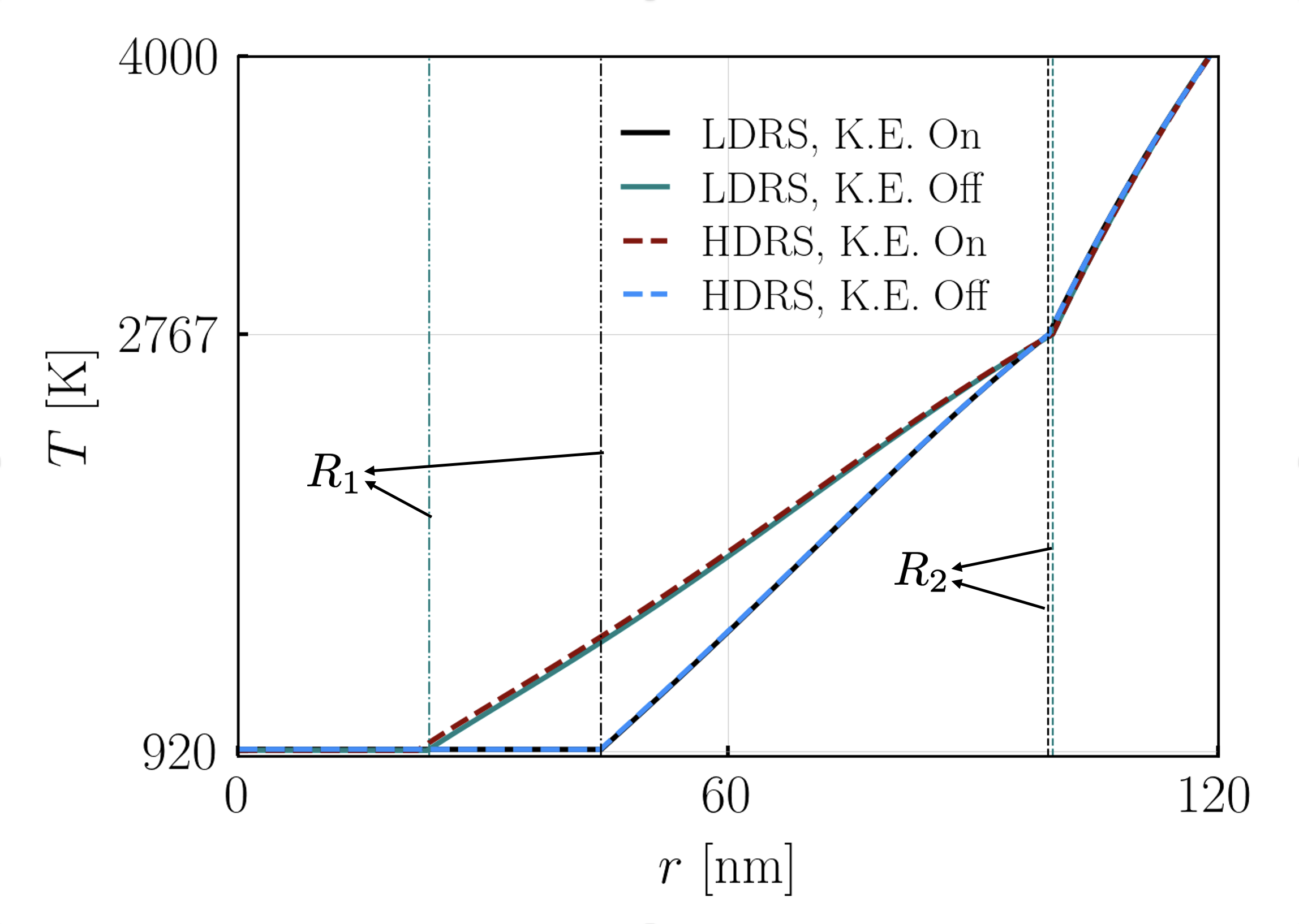}
        \label{fig_TempComparison3P}
        }
	\end{center}
\caption{Time evolution of interface positions and temperature distribution within the domain for a 100 nm Al-like particle, with and without considering the kinetic energy terms in the Stefan conditions. The vapor phase density is taken to be $\rhov = 500~\mathrm{kg/m^3}$. Time evolution of \subref{fig_R1} the melt front, \subref{fig_R2} the boiling front, and the  \subref{fig_Rb} outer free boundary. \subref{fig_TempComparison3P} Temperature distribution within the particle at $\that = 0.1$.} \label{fig_3PKE}
\end{figure}

To establish grid convergence for the three phase problem, we conducted a grid convergence study utilizing both the LDRS and HDRS. The  vapor density is taken as $\rhov = 500$ kg/m$^3$. For the small-time LDRS, we conduct simulations on three grids of size $N = \{ 1677,3353,6704\}$ cells with a fixed CFL number of $c = 0.005$. An additional simulation was conducted using a very fine grid of size $N = 13408$, which served as the reference solution. Similarly, to study the grid convergence of the numerical solution initialized with the small-time HDRS, we conducted simulations on five grids of size $N = \{ 950,1898,2846,4743,5691\}$. The reference solution was obtained from an additional simulation conducted on grid size $N = 7588$. The CFL number was fixed at $c = 0.001$. The $L^2$-norm of error between the numerical solution and the reference solution is calculated for the interface positions (melting and boiling fronts) and the domain temperature. The $L^2$ error for a quantity $\psi$ is defined as the root mean squared error (RMSE) of the vector $||\mathcal{E}_\psi||_{\rm RMSE} = ||\psi_{\rm reference} - \psi_{\rm numerical}||_2 / \sqrt{\mathcal{S}}$. Here, $\mathcal{S}$ represents the size of the vector $\mathcal{E}_\psi$. Figs.~\ref{fig_IntfConv3P},~\ref{fig_IntfR2Conv3P} and~\ref{fig_TempConv3P} plot the actual error data as filled markers and the best fit lines to determine the convergence rate of the error as a function of grid size $N$.  It is observed that the method converges with better than second-order spatiotemporal accuracy for the three-phase Stefan problem. Results from LDRS and HDRS initialized simulations for $\rhov = 500$ kg/m$^3$ case are presented in Fig.~\ref{fig_3PKE}. These correspond to the finest grid simulations. Most of the phase change dynamics (domain temperature and interface position) are not too sensitive to the initial solution, except that the initial dynamics of the boiling front $\Rhattwo$ exhibit a slight overshoot. LDRS is more appropriate for relatively large vapor densities.


\begin{figure}[]\label{fig_3PKE_Rv23}
	\begin{center}
        \subfigure[]{\includegraphics[scale=0.13]{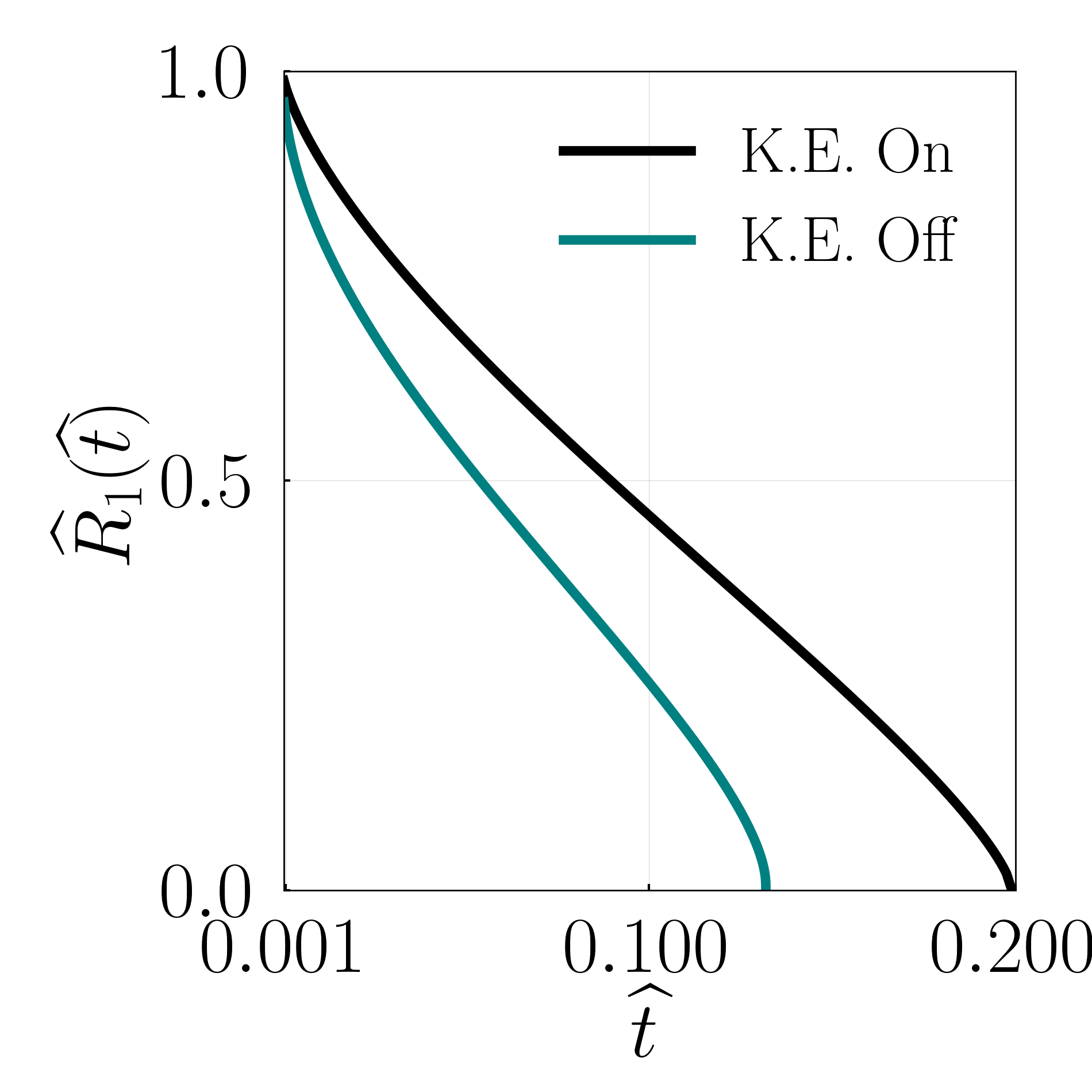}
        \label{fig_R1_Rv23}
        }
        \subfigure[]{\includegraphics[scale=0.13]{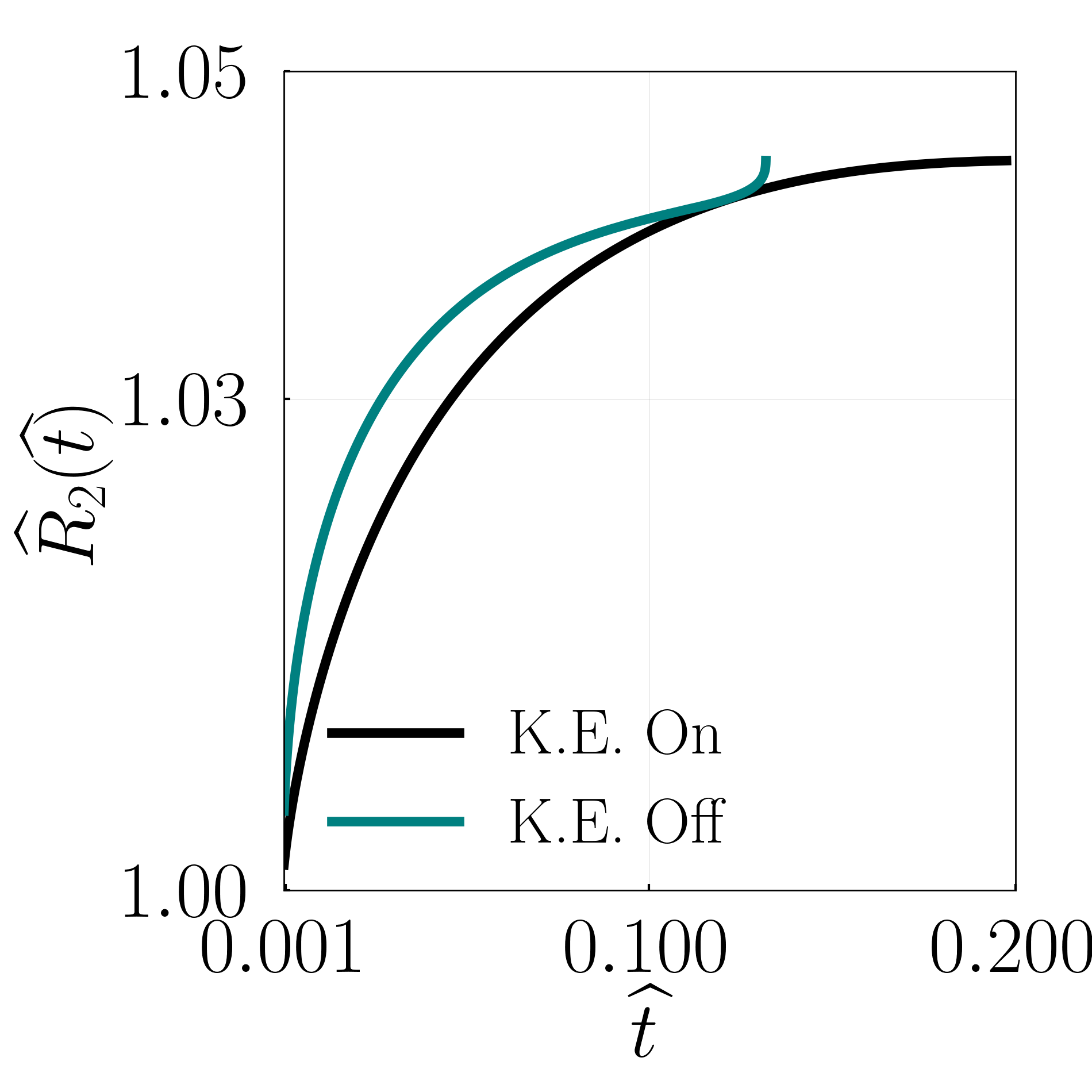}
        \label{fig_R2_Rv23}
        }
        \subfigure[]{\includegraphics[scale=0.13]{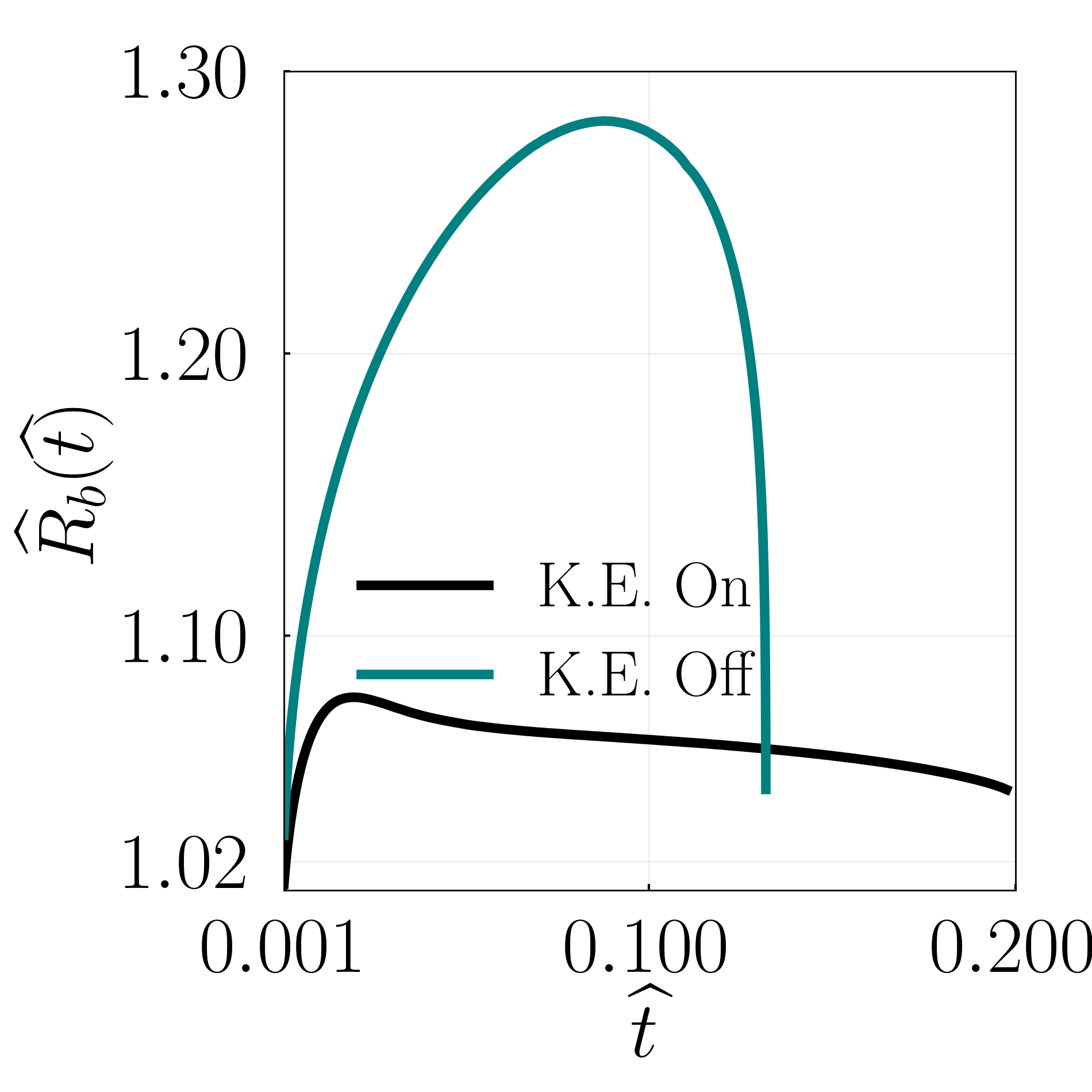}
        \label{fig_Rb_Rv23}
        }
        \subfigure[]{\includegraphics[scale=0.24]{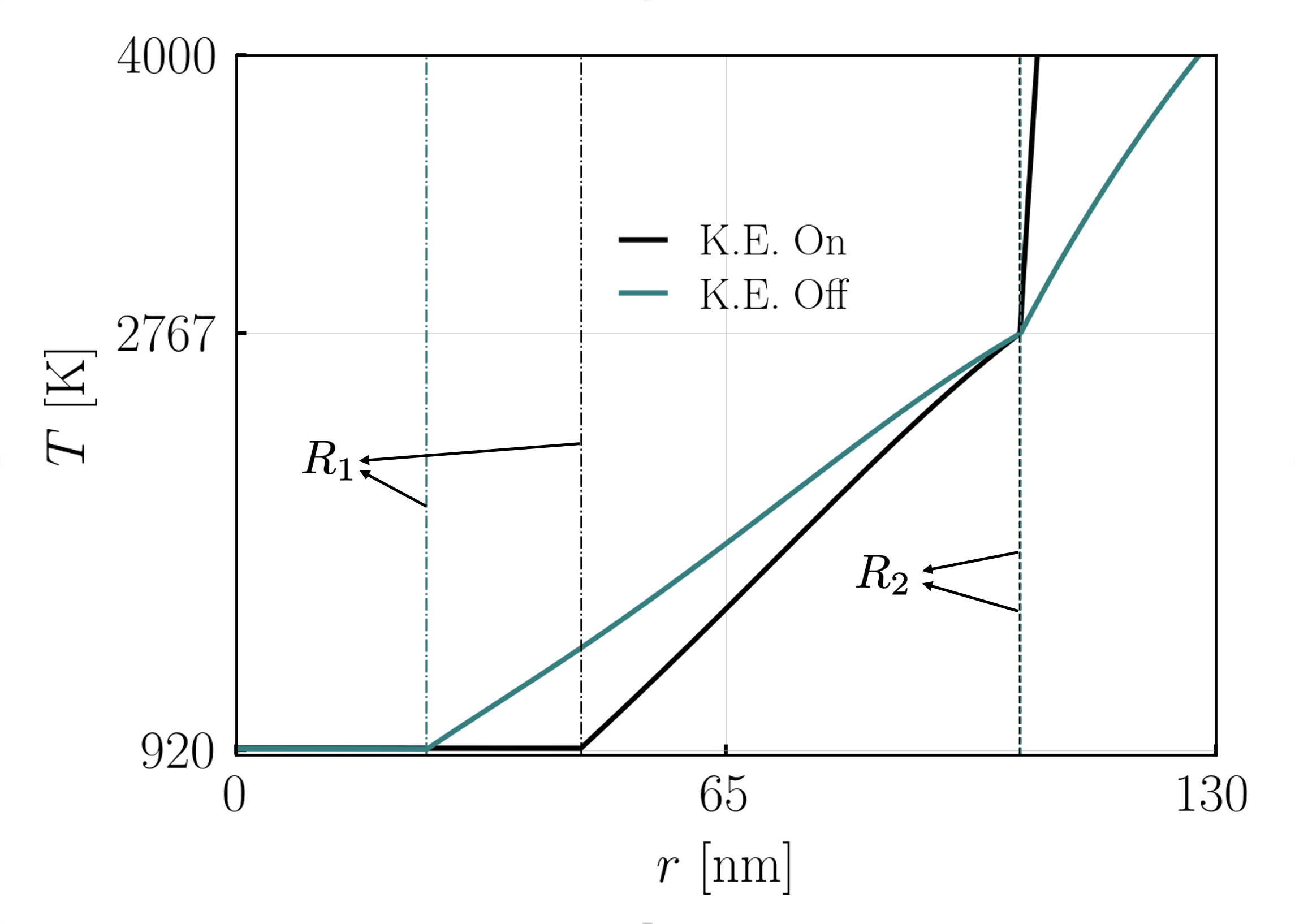}
        \label{fig_TempComparisonRv23}
        }
	\end{center}
\caption{Time evolution of interface positions and temperature distribution within the domain for a 100 nm Al-like particle, with and without considering the kinetic energy terms in the Stefan conditions. The vapor phase density is taken to be $\rhov = 23~\mathrm{kg/m^3}$. Time evolution of \subref{fig_R1_Rv23} the melt front, \subref{fig_R2_Rv23} the boiling front, and \subref{fig_Rb_Rv23} the outer free boundary. \subref{fig_TempComparisonRv23} Temperature distribution within the particle at $\that = 0.1$.} 
\end{figure}

We also analyze the effect of neglecting ($\deltam = \deltav = 0$) and including ($\deltam \neq 0$, $\deltav \neq 0$) the kinetic energy (K.E.) terms in the Stefan conditions for the three-phase spherical Stefan problem. Fig.~\ref{fig_R1} shows that including the kinetic energy terms substantially reduces the melt-front speed and increases the melt time from $\that\approx0.13$ to $\that\approx0.19$, which is an increase of roughly $46\%$. The liquid-vapor interface $\Rhattwo$ in Fig.~\ref{fig_R2} exhibits a sensitivity to kinetic energy terms. The ``K.E. Off'' case shows a more non-monotone trajectory including intervals where $\Rhattwo$ moves outwards as dictated by the particle's mass conservation. The free outer boundary $\Rhatb$ in Fig.~\ref{fig_Rb} expands significantly in both cases. We notice a larger outer boundary radius $\Rhatb$ in the case with ``K.E. On'' by approximately $5\%$. Additionally, Fig.~\ref{fig_TempComparison3P} shows the representative temperature profiles at a non-dimensional time $\that=0.1$ for the two cases. A key difference is the location of the interfaces: with ``K.E. On'' the melt front remains at a larger radius, whereas we do not observe a significant difference in the location of the boiling front. For the given thermophysical and material properties, we notice that the melt front shows a greater sensitivity towards kinetic energy than the boiling front.

Next we consider the low vapor density case of $\rhov = 23$ kg/m$^3$. The remainder of aluminum's thermophysical properties are taken from Table~\ref{tab_thermophys_properties_Al}. To initialize the simulation, we use the small-time HDRS. LDRS predicts nonphysical interface positions that result in negative liquid and vapor layer thicknesses. For this case we adopt an adaptive time stepping scheme that adjusts the step size according to the maximum velocity in the domain. The CFL number is fixed at $c = 0.01$. Fig.~\ref{fig_R1_Rv23} illustrates that including the kinetic energy terms substantially reduces the melt-front speed and increases the melt time from $\that\approx0.13$ to $\that\approx0.2$, an increase of almost $50\%$. Fig.~\ref{fig_R2_Rv23} shows that $\Rhattwo$ is less sensitive to kinetic energy terms. A significant difference can be observed in the outer free boundary location, $\Rhatb$, in Fig.~\ref{fig_Rb_Rv23}. When the kinetic energy terms are ignored, the domain expands and contracts rapidly at the beginning and end of the simulation, respectively. However, the final radius of the outer boundary remains the same. Fig.~\ref{fig_TempComparisonRv23} shows the representative temperature profiles in the domain at non-dimensional time $\that=0.1$ for both cases. There is a significant difference in melt front location, whereas boiling front location is minimally affected.

\begin{figure}[]
	\begin{center}
        \subfigure[$R_0 = 1\;\mu$m]{\includegraphics[scale=0.27]{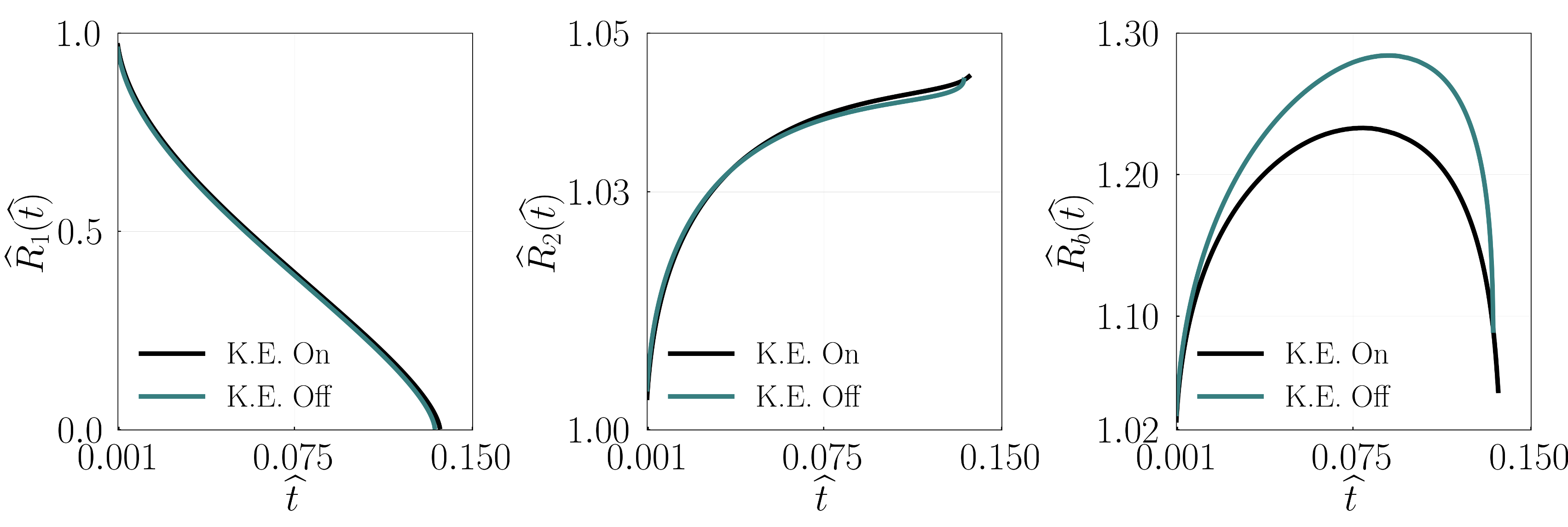}
        \label{fig_1micron}
        }
        \subfigure[$R_0 = 10\; \mu$m]{\includegraphics[scale=0.27]{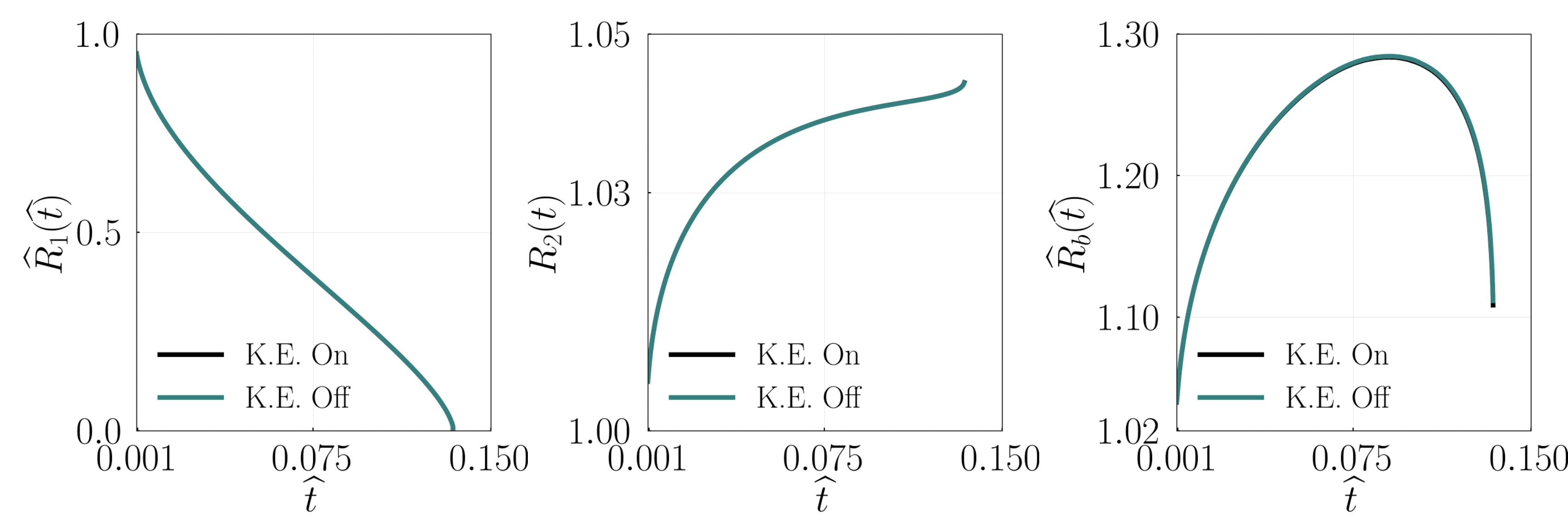}
        \label{fig_10micron}
        }
	\end{center}
\caption{Comparison of the time evolution of $\Rhatone$,  $\Rhattwo$ and $\Rhatb$ including and neglecting the kinetic energy terms in the Stefan conditions for large particles: \subref{fig_1micron} $R_0 = 1~\mu\mathrm{m}$ and \subref{fig_10micron} $R_0 = 10~\mu\mathrm{m}$. The vapor phase density is taken to be $\rhov = 23~\mathrm{kg/m^3}$.} \label{fig_3PLarge}
\end{figure}

We also simulate the melting and boiling of large particles of sizes 1 and 10$~\mu\mathrm{m}$. Fig.~\ref{fig_3PLarge} illustrates the time evolution of the melt and boiling fronts, and the outer free boundary with and without retaining the kinetic energy terms in the Stefan conditions. As can be observed in the figure, for large particles, there is a negligible effect of kinetic energy on phase change dynamics.

\section{Conclusions and discussion}

In this work, we presented a mathematical model for two- and three-phase spherical Stefan problems that consistently accounts for variation in thermophysical properties, including density between various phases of a phase change material. The Stefan conditions consider a jump in kinetic energy across the phase change interface, which is typically ignored in the literature. We also presented a fixed-grid, sharp-interface method for simulating two- and three-phase spherical Stefan problems which was demonstrated to be second-order in spatiotemporal accuracy. Small-time analytical solutions were derived to initialize the simulation. Large and small density contrasts between the phases are considered to derive the appropriate small-time solutions. The numerical method and the small-time analysis are validated against the results reported in the literature for the two-phase case before utilizing them to solve the three-phase problems. For the three-phase Stefan problem, we quantified the influence of kinetic energy terms ($\deltam$ and $\deltav$) on the time evolution of melt and boiling fronts, as well as on the outer free boundary of a nanoparticle. It is found that including kinetic energy terms predicts a longer melting time for the particle than when it is not. The melt front dynamics exhibited greater sensitivity to kinetic energy terms than the boiling front. For the larger particles of size micrometers and above, the kinetic energy terms do not affect the phase change dynamics. In summary, the present study tackles a rather unexplored, but physically relevant heat and mass transfer problem, which is the three-phase version of the classical Stefan problem.

\section*{Acknowledgements}
A.P.S.B~acknowledges support of NSF award CBET CAREER 2234387. 

\begin{appendix}

\section{Additional grid convergence results for the two-phase Stefan problem}\label{sec_2P_Stefan_Convergence}

\begin{figure}[]
	\begin{center}
        \subfigure[Melt front position]{\includegraphics[scale=0.14]{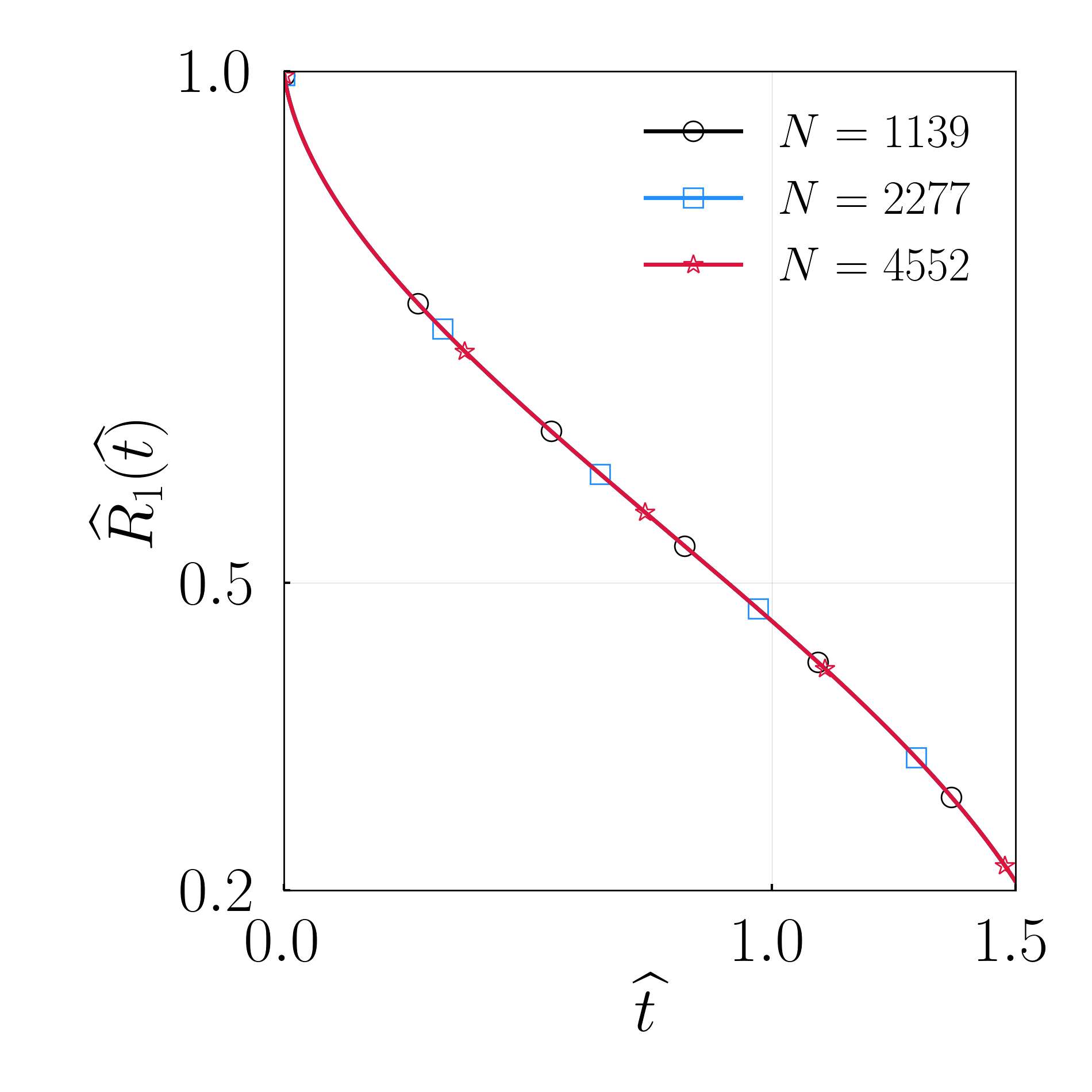}
        \label{fig_Intf2P}
        }
        \subfigure[Temperature at $\that = 1.36$]{\includegraphics[scale=0.14]{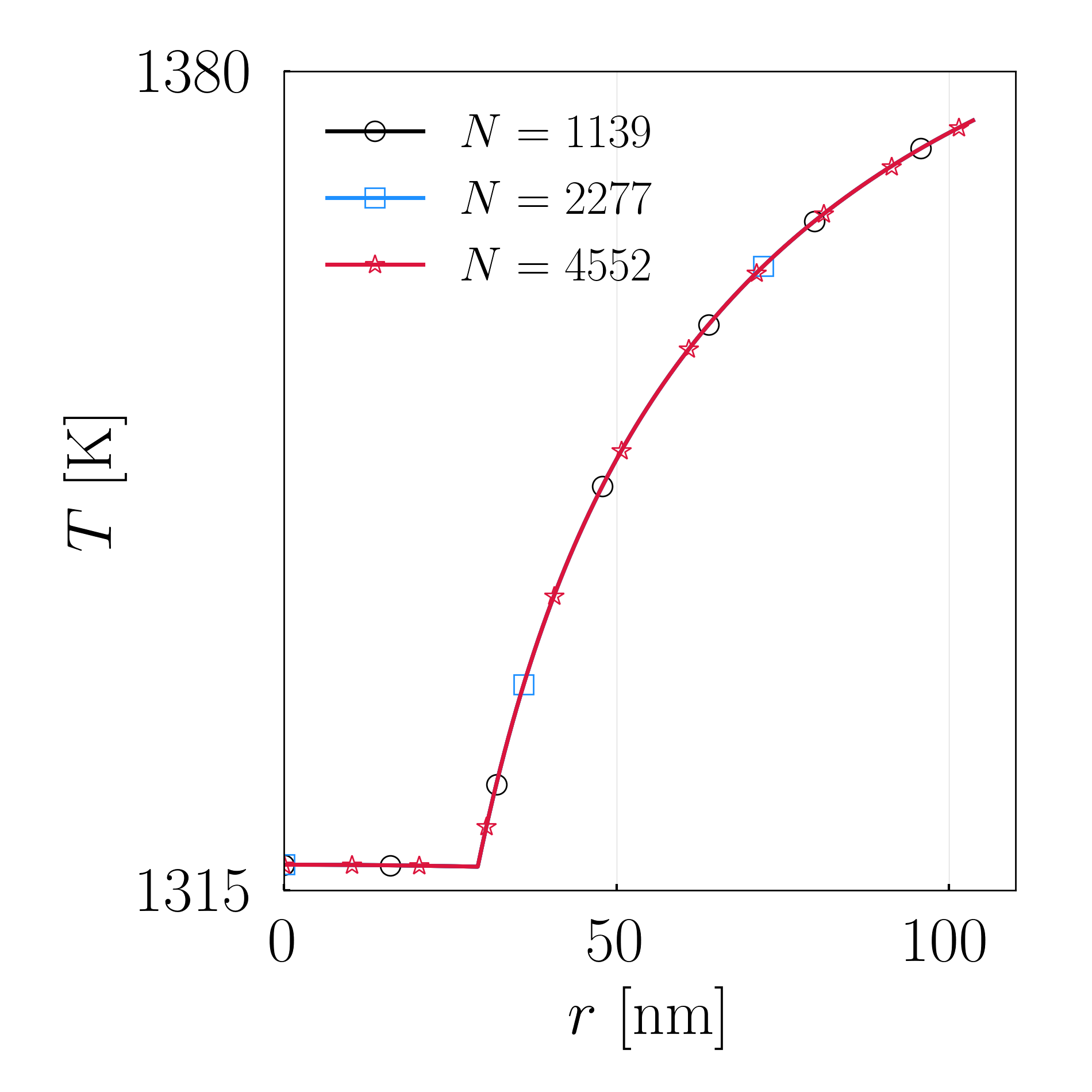}
        \label{fig_Temp2P}
        }
	\end{center}
\caption{Grid convergence of \subref{fig_Intf2P} melt front position and \subref{fig_Temp2P} temperature in the nanoparticle for the two-phase Stefan problem presented in Sec.~\ref{sec_2P_Stefan_results} for Stefan number $\betam = 10$ and $\rrhosl \ne 1$.} \label{fig_2PGridConvg}
\end{figure}
 
 Here we conduct a grid convergence test for the two-phase Stefan problem considering $\betam = 10$, $\rhos = 19300$ kg/m$^3$,   and $\rhol = 17300$ kg/m$^3$.  The simulations are conducted on three grids of size $N = \{ 1139,2277,4552\}$. The CFL number is fixed at $c = 0.005$ for all grids. The simulations are run from $\that= 0.001$ till $\that=1.5$. Fig.~\ref{fig_2PGridConvg} illustrates that the numerical results (melt front position and temperature in the domain) are essentially grid independent.
\end{appendix}
 
\bibliography{Ref.bib}

\begin{thebibliography}{10}
\expandafter\ifx\csname url\endcsname\relax
  \def\url#1{\texttt{#1}}\fi
\expandafter\ifx\csname urlprefix\endcsname\relax\def\urlprefix{URL }\fi
\expandafter\ifx\csname href\endcsname\relax
  \def\href#1#2{#2} \def\path#1{#1}\fi

\bibitem{Mehran2025analyticalmsnbc}
M.~Soleimani, K.~Koponen, N.~Tilton, A.~P.~S. Bhalla, Analytical and numerical
  solutions to the three-phase stefan problem with simultaneous occurrences of
  melting, solidification, boiling, and condensation phenomena, ASME Journal of
  Heat and Mass Transfer 148~(1) (2026) 012401.

\bibitem{font2015nanoparticle}
F.~Font, T.~G. Myers, S.~L. Mitchell, A mathematical model for nanoparticle
  melting with density change, Microfluidics and Nanofluidics 18 (2015)
  233--243.

\bibitem{vuik1993some}
C.~Vuik, {Some historical notes about the Stefan problem}, Tech. rep., Delft
  University of Technology, Faculty of Technical Mathematics and Informatics
  (1993).

\bibitem{rubinvsteuin2000stefan}
L.~I. Rubinvsteuin, {The Stefan problem}, Vol.~8, American Mathematical Soc.,
  2000.

\bibitem{hahn2012heat}
D.~W. Hahn, M.~N. {\"O}zisik, Heat conduction, John Wiley \& Sons, 2012.

\bibitem{alexiades2018mathematical}
V.~Alexiades, A.~D. Solomon, Mathematical modeling of melting and freezing
  processes, Routledge, 2018.

\bibitem{mccue2009micro}
S.~W. McCue, B.~Wu, J.~M. Hill, Micro/nanoparticle melting with spherical
  symmetry and surface tension, IMA journal of applied mathematics 74~(3)
  (2009) 439--457.

\bibitem{mccue2008classical}
S.~W. McCue, B.~Wu, J.~M. Hill, Classical two-phase stefan problem for spheres,
  Proceedings of the Royal Society A: Mathematical, Physical and Engineering
  Sciences 464~(2096) (2008) 2055--2076.

\bibitem{font2013spherically}
F.~Font, T.~Myers, Spherically symmetric nanoparticle melting with a variable
  phase change temperature, Journal of nanoparticle research 15~(12) (2013)
  2086.

\bibitem{wu2009nanoparticle}
B.~Wu, P.~Tillman, S.~W. McCue, J.~M. Hill, Nanoparticle melting as a stefan
  moving boundary problem, Journal of Nanoscience and Nanotechnology 9~(2)
  (2009) 885--888.

\bibitem{myers2020stefan_varprops}
T.~G. Myers, M.~G. Hennessy, M.~Calvo-Schwarzw{\"a}lder, The stefan problem
  with variable thermophysical properties and phase change temperature,
  International Journal of Heat and Mass Transfer 149 (2020) 118975.
\newblock \href {https://doi.org/10.1016/j.ijheatmasstransfer.2019.118975}
  {\path{doi:10.1016/j.ijheatmasstransfer.2019.118975}}.

\bibitem{koponen2025direct}
K.~Koponen, A.~P.~S. Bhalla, B.~Sprinkle, N.~Wu, N.~Tilton, A direct forcing,
  immersed boundary method for conjugate heat transport, Journal of
  Computational Physics (2025) 114135.

\bibitem{nangia2019robust}
N.~Nangia, B.~E. Griffith, N.~A. Patankar, A.~P.~S. Bhalla, A robust
  incompressible navier-stokes solver for high density ratio multiphase flows,
  Journal of Computational Physics 390 (2019) 548--594.

\bibitem{buffat1976size}
P.~Buffat, J.-P. Borel, Size effect on the melting temperature of gold
  particles, Physical Review A 13~(6) (1976) 2287--2297.

\bibitem{thirumalaisamy2023low}
R.~Thirumalaisamy, A.~P.~S. Bhalla, A low mach enthalpy method to model
  non-isothermal gas--liquid--solid flows with melting and solidification,
  International Journal of Multiphase Flow 169 (2023) 104605.

\bibitem{doble2007perry}
M.~Doble, Perry’s chemical engineers’ handbook, McGraw-Hil, New York, US
  (2007).

\bibitem{hatch1984aluminium}
J.~E. Hatch, Aluminium: Properties and physical metallurgy, by asm, Metals
  Park, OH 135 (1984).

\bibitem{desai1987thermodynamic}
P.~Desai, Thermodynamic properties of aluminum, International journal of
  thermophysics 8 (1987) 621--638.

\end{thebibliography}
\end{document}